\documentclass[twocolumn]{aastex62}

\usepackage[normalem]{ulem}
\useunder{\uline}{\ul}{}
\usepackage{lineno}
\usepackage{color}
\usepackage{subfigure}
\usepackage{amsthm}
\usepackage{amsmath}
\usepackage{hyperref}

\revised{\today}

\shorttitle{}
\shortauthors{}

\begin{document}

\title{Occurrence of gravitational collapse in the accreting neutron stars of binary-driven hypernovae}

\author{L.~M.~Becerra}
\affiliation{\textit{GIRG}, Escuela de F\'isica, Universidad Industrial de Santander, 680002 Bucaramanga, Colombia}
\affiliation{ICRANet, Piazza della Repubblica 10, I-65122 Pescara, Italy}

\author{F.~Cipolletta}
\affiliation{Barcelona Supercomputing Center, Plaça Eusebi Güell 1--3, I--08034 Barcelona, Spain}

\author{C.~L.~Fryer}
\affiliation{CCS-2, Los Alamos National Laboratory, Los Alamos, NM 87545}

\author{Débora P. Menezes}
\affiliation{Departamento de F\'isica - CFM, Universidade Federal de Santa Catarina, Florian\'opolis/SC, CEP 88.040--900, Brazil}

\author{Constança Providência}
\affiliation{CFisUC, Department of Physics, University of Coimbra, 3004--516 Coimbra, Portugal}

\author{J.~A.~Rueda}
\affiliation{ICRANet, P.zza della Repubblica 10, 65122 Pescara, Italy}
\affiliation{ICRA, Dipartimento di Fisica, Sapienza Universit\`a di Roma, P.le Aldo Moro 5, 00185 Rome, Italy}
\affiliation{ICRANet-Ferrara, Dipartimento di Fisica e Scienze della Terra, Universit\`a degli Studi di Ferrara, Via Saragat 1, I--44122 Ferrara, Italy}
\affiliation{Dipartimento di Fisica e Scienze della Terra, Universit\`a degli Studi di Ferrara, Via Saragat 1, I--44122 Ferrara, Italy}
\affiliation{INAF, Istituto di Astrofisica e Planetologia Spaziali, Via Fosso del Cavaliere 100, 00133 Rome, Italy}

\author{R.~Ruffini}
\affiliation{ICRANet, Piazza della Repubblica 10, I-65122 Pescara, Italy}
\affiliation{ICRA, Dipartimento di Fisica, Sapienza Universit\`a di Roma, P.le Aldo Moro 5, 00185 Rome, Italy}
\affiliation{INAF,Viale del Parco Mellini 84, 00136 Rome, Italy}

\email{laura.becerra7@correo.uis.edu.co}  

\begin{abstract}
The binary-driven hypernova (BdHN) model proposes long gamma-ray bursts (GRBs) originate in binaries composed of a carbon-oxygen (CO) star and a neutron star (NS) companion. The CO core collapse generates a newborn NS and a supernova that triggers the GRB by accreting onto the NSs, rapidly transferring mass and angular momentum to them. {This article aims to determine the conditions under which a black hole (BH) forms from NS collapse induced by the accretion and the impact on the GRB observational properties and taxonomy.} We perform three-dimensional, smoothed-particle-hydrodynamics simulations of BdHNe using up-to-date NS nuclear equations of state (EOS), with and without hyperons, and calculate the structure evolution in full general relativity. We assess the binary parameters leading either NS {in the binary} to the critical mass for gravitational collapse into a BH and its occurrence time, $t_{\rm col}$. We include a non-zero angular momentum of the NSs and find that $t_{\rm col}$ ranges from a few tens of seconds to hours for decreasing NS initial angular momentum values. BdHNe I are the most compact (about five minutes orbital period), promptly form a BH and release $\gtrsim 10^{52}$ erg. They form NS-BH binaries with tens of kyr merger timescale by gravitational-wave emission. BdHNe II and III do not form BHs, release $\sim 10^{50}$--$10^{52}$ erg and $\lesssim 10^{50}$ erg. {They form NS-NS binaries with a range of merger timescales larger than for NS-BH binaries.}. In some compact BdHNe II, either NS can become supramassive, i.e., above the critical mass of a non-rotating NS. Magnetic braking by a $10^{13}$ G field can delay BH formation, leading to BH-BH or NS-BH of tens of kyr merger timescale.
\end{abstract}


%%%%%%%%%%%%%%%%%%%%%%%%%%%%%%%%%%%%%%%%%%%%%%%%%%
%%%%%%%%%%%%%%%%% BODY OF PAPER %%%%%%%%%%%%%%%%%%

%%%%%%%%%%%%%%%%%%%%%%%%%%%%%%%%%%%%%%%%%%%%%%%%%%%%%%%%%%%
%%%%%%%%%%%%%%%%%%%%%%%%%%%%%%%%%%%%%%%%%%%%%%%%%%%%%%%%%%%
\section{Introduction}\label{sec:1}
%%%%%%%%%%%%%%%%%%%%%%%%%%%%%%%%%%%%%%%%%%%%%%%%%%%%%%%%%%%
%%%%%%%%%%%%%%%%%%%%%%%%%%%%%%%%%%%%%%%%%%%%%%%%%%%%%%%%%%%

Understanding the physical and astrophysical conditions under which a neutron star (NS) can reach the point of gravitational collapse into a black hole (BH) is essential for understanding the final stages of binary stellar evolution, which are associated with the most energetic and powerful cataclysms in the Universe: {gamma-ray bursts} (GRBs). 

In a few seconds, a GRB can {produce} a gamma-ray luminosity comparable to the luminosity of all stars in the observable Universe, which makes a GRB detectable to cosmological redshifts $z\sim 10$, close to the dawn of galaxy and stellar formation. Observationally, they are classified as short or long depending on whether $T_{90}$ is shorter or longer than $2$ s. The time $T_{90}$ is the time interval in the observer frame where $90\%$ of the isotropic energy in gamma-rays ($E_{\rm iso}$) is released \citep{1981Ap&SS..80....3M, 1992grbo.book..161K, 1992AIPC..265..304D, 1993ApJ...413L.101K, 1998ApJ...497L..21T}. This article focuses on assessing the {BH formation in long GRBs within the binary-driven hypernova (BdHN) scenario (details below)}.

{Despite a few peculiar exceptions of short bursts that shows hybrid properties of long GRBs, pointing to alternative scenarios \citep[e.g., discussion in][]{2021NatAs...5..911Z,2018JCAP...10..006R}, NS binary (NS-NS) and NS-BH mergers are widely accepted as the progenitors of the short GRBs \citep{1986ApJ...308L..47G, 1986ApJ...308L..43P, 1989Natur.340..126E, 1991ApJ...379L..17N}. This view has gained additional attention by the proposed first electromagnetic counterpart associated with a gravitational-wave event, i.e., GW170817 and GRB 170817A \citep{2017PhRvL.119p1101A}.
}

For long GRBs, the model \citep[see, e.g.,][for reviews]{2002ARA26A..40..137M,2004RvMP...76.1143P} based on a relativistic jet, a \textit{fireball} of an optically thick $e^-e^+$-photon-baryon plasma \citep{1978MNRAS.183..359C, 1986ApJ...308L..43P, 1986ApJ...308L..47G, 1991ApJ...379L..17N, 1992ApJ...395L..83N}, with bulk Lorentz factor $\Gamma \sim 10^2$--$10^3$ \citep{1990ApJ...365L..55S, 1992MNRAS.258P..41R, 1993MNRAS.263..861P, 1993ApJ...415..181M, 1994ApJ...424L.131M}, powered by a massive disk accreting onto a BH, became the traditional GRB model. The formation of such a BH-massive disk structure has been theorized by the \textit{collapsar} model \citep{1993ApJ...405..273W, 1999ApJ...524..262M}, i.e., the core-collapse of a single, massive, fast-rotating star. In parallel, it has also been explored as GRB central engine, a highly-magnetized, newborn NS \citep{1992Natur.357..472U,2000ApJ...537..810W,2011MNRAS.413.2031M}, currently referred to as \textit{millisecond magnetar} scenario. This model has gained attention in the literature mainly in the explanation of low to moderate luminosity GRBs \citep[see, e.g.,][and references therein]{2004ApJ...611..380T,2007MNRAS.380.1541B,2007ApJ...659..561M,2017MNRAS.472.3058B,2018ApJ...857...95M,2023ApJ...949L..32D,2023ApJ...952..156S}; see also \citet{2018pgrb.book.....Z} for a recent review of the traditional GRB model.

Meanwhile, GRB astronomy {has made significant advances} that challenge the traditional picture of long GRBs and evidence the need to explore alternatives. In particular, there is mounting evidence of the necessity of accounting for binary progenitors, as discussed below.

First, we recall the X-ray afterglow discovery by the BeppoSAX satellite \citep{1997Natur.387..783C} and the confirmation of the GRB cosmological nature \citep{1997Natur.387..878M}. The accurate source localization provided by BeppoSAX allowed the optical follow-up by ground-based telescopes. This led to one of the most significant discoveries: the observation of long GRBs in coincidence with type Ic supernovae (SNe). The first GRB-SN association was GRB 980425-SN 1998bw \citep{1998Natur.395..670G}. The number of associations has since then increased thanks to the optical afterglow follow-up by the Neil Gehrels Swift Observatory \citep{2005SSRv..120..143B,2005SSRv..120..165B,2005SSRv..120...95R}, leading to about twenty {robust} cases {with spectroscopic coverage as} of today \citep{2006ARA&A..44..507W, 2011IJMPD..20.1745D, 2012grb..book..169H,2017AdAst2017E...5C,2023ApJ...955...93A}. {Further potential GRB-SN associations have been claimed, although they are uncertain owing to the lack of spectral verification \citep[see, e.g., discussion in][]{2017AdAst2017E...5C,2022ApJ...938...41D}.}.

Interestingly, {all SNe associated with long GRBs show} similar luminosity and time of occurrence (from the GRB trigger). In contrast, the associated GRBs show energy releases that span nearly seven orders of magnitude \citep[see][for details]{2023ApJ...955...93A}. It seems challenging to reconcile this observational result with a model based on a single massive stellar collapse.

The association of GRB-SN systems to massive star explosions has been set statistically \citep{2006Natur.441..463F,2008ApJ...689..358R,2008ApJ...687.1201K} and from the modeling of photometric and spectroscopic observations of the optical emission of the GRB-associated SNe \citep[see][for specific examples]{2006ARA&A..44..507W, 2006ApJ...640..854M, 2003omeg.conf..223N, 2003ApJ...599L..95M, 2006ApJ...645.1323M, 2009ApJ...700.1680T, Bufano...2012ApJ...753...67B, 2019MNRAS.487.5824A}. Thus, further constraints and drawbacks of a single-star collapse model for long GRBs arise from observational evidence pointing out the ubiquitous role of binaries in the stellar evolution of massive stars.

Observations show that most massive stars belong to binaries \citep{2007ApJ...670..747K,2012Sci...337..444S}. {For a more recent analysis, we refer to \citet{2020ApJ...900..118N}. The ubiquitousness of massive binaries across the universe has also caught the attention of the James Webb Space Telescope (JWST) with the recent observations of Mothra, a likely binary of two supergiants at redshift $z = 2.091$ \citep{2023A&A...679A..31D}. A new observational exploration era has started, looking for binary companions of SN explosions in binaries \citep[see, e.g.,][and references therein]{2021MNRAS.505.2485O,2022ApJ...929L..15F,2023ApJ...956L..31M,2023ApJ...949..121C,2024Natur.625..253C}.}

This makes particular contact with GRBs since they are associated with SNe of type Ic. These SNe lack hydrogen (H) and helium (He), and most SN Ic models acknowledge binary interactions as the most effective mechanism to get rid of the H and He layers of the pre-SN star \citep[see, e.g.,][]{1988PhR...163...13N, 1994ApJ...437L.115I, 2007PASP..119.1211F, 2010ApJ...725..940Y, 2011MNRAS.415..773S, 2015PASA...32...15Y, 2015ApJ...809..131K}.

From the theoretical side, it appears an extreme assumption that the gravitational collapse of a single massive star forms a collapsar, a jetted fireball, and an SN explosion{, although} some proposals have been made to mitigate this difficulty \citep[see, e.g.,][]{1999ApJ...524..262M,2005ApJ...629..341K, 2012ApJ...744..103M, 2012ApJ...750..163L, 2015ApJ...810..101F}. Theoretical models fail to produce the fast rotation to produce a GRB jet and an SN-like explosion from a single-stellar collapse \citep{2007PASP..119.1211F}. Moreover, stellar evolution suggests that the angular momenta in stellar cores of single stars are even less than what earlier models obtained \citep{2015ApJ...810..101F}. Binary interactions are likely required to produce the high angular momenta needed for the BH-accretion disk mechanism of the traditional GRB model. However, extreme fine-tuning appears necessary to reach the requested {physical} conditions, challenging to reproduce the GRB density rates \citep[see, e.g.,][]{2022MNRAS.511.3951F}.

{\citet{2009ARA&A..47...63S,2009MNRAS.395.1409S,2015PASA...32...16S} analyzed archival images of the locations of past SNe to pinpoint their pre-SN stars. The result has been that the inferred zero-age main-sequence (ZAMS) progenitors masses are $\lesssim  18 M_\odot$. These observations agree with standard stellar evolution, i.e., ZAMS stars $\lesssim 25 M_\odot$ evolve to core-collapse SNe, forming NSs. Direct BH formation occur for ZAMS above $20$--$25 M_\odot$, without SN \citep[e.g.,][]{1999ApJ...522..413F, 2003ApJ...591..288H, 2022ApJ...938...69P}. Instead, they are at odds with the traditional GRB model, which requests a single-star core collapse to produce a BH and an SN.}

All the above facts on GRB-SNe {strongly suggest} that most GRBs, if not all, should occur in binaries. In their pioneering work, this possibility was envisaged by \citet{1999ApJ...526..152F}. The alternative BdHN model has been developed {complementing and joining the SN and binary evolution community results, filling} the gap between the increasing observational evidence of the role of binaries in the stellar evolution of massive stars and theoretical modeling of long GRBs. Specifically, the BdHN model proposes the GRB event occurs in a binary composed of a carbon-oxygen star (CO) and an NS companion. We refer the reader to \citep{2012ApJ...758L...7R, 2012A&A...548L...5I, 2014ApJ...793L..36F, 2015PhRvL.115w1102F, 2015ApJ...812..100B, 2016ApJ...833..107B, 2019ApJ...871...14B, 2022PhRvD.106h3002B} for theoretical details on the model. The core of the CO star collapses, generating a newborn NS (hereafter, $\nu$NS) and the SN. The latter triggers the GRB event, which shows up to seven emission episodes in the most energetic sources, associated with specific physical processes that occur in a BdHN, scrutinized in recent years \citep[see][and references therein]{2023ApJ...955...93A}. Numerical three-dimensional (3D) smoothed-particle-hydrodynamics (SPH) simulations of the SN explosion in the CO-NS binary \citep{2019ApJ...871...14B,2022PhRvD.106h3002B} show the CO-NS fates explain the diversity of GRBs: BdHNe I are the most extreme with energies $10^{52}$--$10^{54}$ erg {with orbital periods of a few minutes}. In these sources, the material ejected in the SN is easily accreted by the NS companion, so it reaches the point of gravitational collapse, forming a rotating BH. In BdHNe II, the orbital period {is of a few tens of minutes} and emit energies $10^{50}$--$10^{52}$ erg. The accretion is lower, so the NS remains stable. The energy separatrix of $10^{52}$ erg of BdHN I and II is set by the energy released when bringing the NS to the critical mass and forming a rotating BH \citep{2016ApJ...832..136R,2018ApJ...859...30R}. The BdHNe III have orbital periods of hours, and the accretion is negligible. They explain GRBs with energies lower than $10^{50}$ erg.

Therefore, the BdHN scenario complements and joins the SN and binary evolution community results, leading to a more comprehensive and global picture of GRB progenitors. {Section \ref{sec:6a} describes the connection between the CO-NS binary and the BdHN I, II, and III features. In this line, a crucial point is the determination of the conditions under which BH formation occurs since it separates BdHNe I from the types II and III}. This is the topic of this work. The article is organized as follows. Section \ref{sec:2} describes the SPH numerical simulations of the GRB-SN event, focusing on determining the accretion rate of material from the SN explosion onto the $\nu$NS and the NS companion. Section \ref{sec:3} details the up-to-date nuclear EOS used in the numerical simulations to describe the NS interiors. Section \ref{sec:4} sets the theoretical framework to determine the evolution of the NS structure during the accretion process and the critical mass limit for gravitational collapse into a rotating BH. Specific results of numerical simulations and the evolution of the NSs and the collapse times are shown in section \ref{sec:5}. {We discuss in section \ref{sec:6} the impact of the results of this work on the BdHN scenario}. Finally, in section \ref{sec:7}, we discuss and draw the main conclusions.

%%%%%%%%%%%%%%%%%%%%%%%%%%%%%%%%%%%%%%%%%%%%%%%%%%%%%%%%%%
%%%%%%%%%%%%%%%%%%%%%%%%%%%%%%%%%%%%%%%%%%%%%%%%%%%%%%%%%%
\section{Simulation of the BdHN early evolution}\label{sec:2}
%%%%%%%%%%%%%%%%%%%%%%%%%%%%%%%%%%%%%%%%%%%%%%%%%%%%%%%%%%
%%%%%%%%%%%%%%%%%%%%%%%%%%%%%%%%%%%%%%%%%%%%%%%%%%%%%%%%%%

%
\begin{figure*}
    \centering
    \includegraphics[width=0.95\hsize,clip]{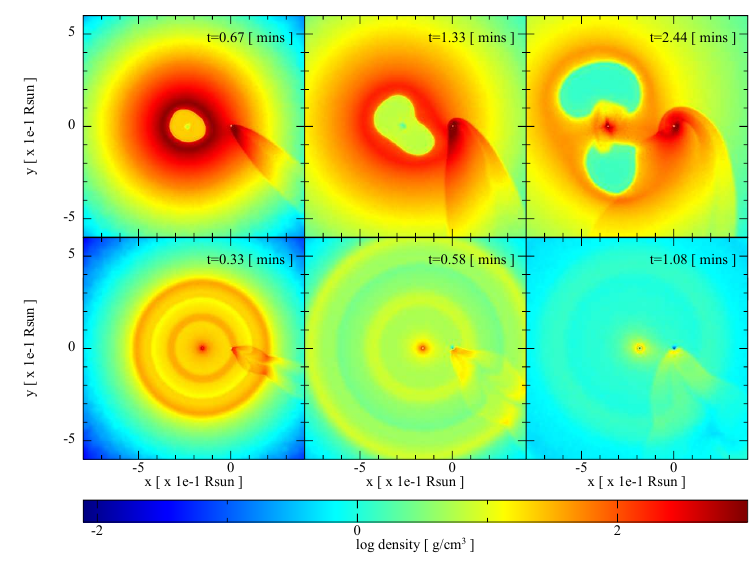}
    \caption{SPH simulations of  BdHNs:  model ``30m1p1eb'' (top) and ``15m1p05e'' (bottom) of Table 2 in \citet{2019ApJ...871...14B}. Top: The binary progenitor comprises a CO star of $\approx 9 M_\odot$, produced by a ZAMS star of $30 M_\odot$, and a $1.9 M_\odot$ NS companion. The orbital period is $\approx 6$ min {and the energy of the SN, $3.26\times 10^{51}$~erg}. Bottom: The binary progenitor is a CO star of $\approx 3 M_\odot$, produced by a ZAMS star of $15 M_\odot$, and a $1.4 M_\odot$ NS companion. The orbital period is $\approx 5$ min {and the energy of the SN, $1.1\times 10^{51}$~erg}.  Each frame corresponds to selected increasing times from left to right with $t = 0$ s the instant of the SN shock breakout. They show the mass density on the equatorial plane.
    The reference system is rotated and translated to align the x-axis with the line joining the binary components. The origin of the reference system is located at the NS companion position. Top: the first frame corresponds to $t = 0.6$ min, showing that the particles entering the NS capture region form a tail behind them. These particles then circularize around the NS, forming a thick disk already visible in the second frame at $t = 1.3$ min. Part of the SN ejecta is also attracted by the $\nu$NS accreting onto it; this is appreciable in the third frame at $t=2.4$ min. Bottom: in all three panels, it can be seen how the material leaves the system almost without being affected by the NS companion.
    This figure has been produced with the SNsplash visualization program \citep{2011ascl.soft03004P}.
    }
    \label{fig:3DSPH}
\end{figure*}

We perform smoothed-particle-hydrodynamics (SPH) simulations with the \textit{SNSPH} code adapted to the binary progenitor of the BdHN presented in \citet{2019ApJ...871...14B}. This Newtonian, three-dimensional (3D) Lagrangian code calculates the evolution of the position, momentum (linear and angular), and thermodynamics (pressure, density, temperature) of the pseudo-particles, which are of mass $m_i$, assigned according to the mass-density distribution of the ejecta. The code estimates the baryonic-mass accretion rate at every time $t = t_0 + \Delta t$, being $t_0$ the simulation's starting time, i.e., the time of the SN explosion set by the time when the SN shock front reaches the CO star surface, as 
\begin{align}
    \dot{M}_{\nu {\rm NS}}^{\rm cap} &= \sum_i m_i \frac{N_{\nu {\rm NS}}^{\rm cap}(\Delta t, m_i)}{\Delta t},\\
    \dot{M}_{\rm NS}^{\rm cap} &= \sum_i m_i \frac{N_{\rm NS}^{\rm cap}(\Delta t, m_i)}{\Delta t}.
\end{align}
where $N_{\nu \rm NS}^{\rm cap}$ and $N_{\rm NS}^{\rm cap}$ are the number of particles gravitationally captured by the $\nu$NS and the NS companion, at the time $t$. The Newtonian scheme suffices for the accretion rate estimate because the size of the gravitational capture region (i.e., the Bondi-Hoyle radius) of the NSs is $100$--$1000$ larger than their Schwarzschild radius \citep{2019ApJ...871...14B}. The gravitational mass and angular momentum of the NSs are calculated in full general relativity by solving, at every time step, the Einstein equations in axial symmetry (see section \ref{sec:4}).

The top panel of Fig. \ref{fig:3DSPH} shows snapshots of the mass density of the SN ejecta in the y-x plane, the binary equatorial plane %, and the z-x plane 
 at different times. In this simulation, the mass of the CO star, just before its collapse, is around $8.89~M_\odot$. This pre-SN configuration is obtained from the thermonuclear evolution of a ZAMS star of $M_{\rm ZAMS}=30~M_\odot$. The NS companion has $1.9~M_\odot$, and the pre-SN orbital period is $5.77$~min, which is the shortest orbital period for the system to avoid Roche-lobe overflow before the SN explosion of the CO core \citep[see, e.g.,][]{2014ApJ...793L..36F}. 

The SPH simulation maps to a 3D-SPH configuration, the 1D core-collapse supernova simulation of \citet{2018ApJ...856...63F}. At this moment, the collapse of the CO star has formed a $\nu$NS of $1.75~M_\odot$, and around $7.14~M_\odot$ is ejected by the SN explosion {of energy $3.26\times 10^{51}$~erg. In the simulation}, the $\nu$NS and NS companion are modeled as point-like masses, interacting only gravitationally with the SN particles and between them. We allow these two point-like particles to increase their mass by accreting other SN particles following the algorithm described in \citet{2019ApJ...871...14B}.

\begin{figure*}
    \centering
   \includegraphics[width=0.49\hsize,clip]{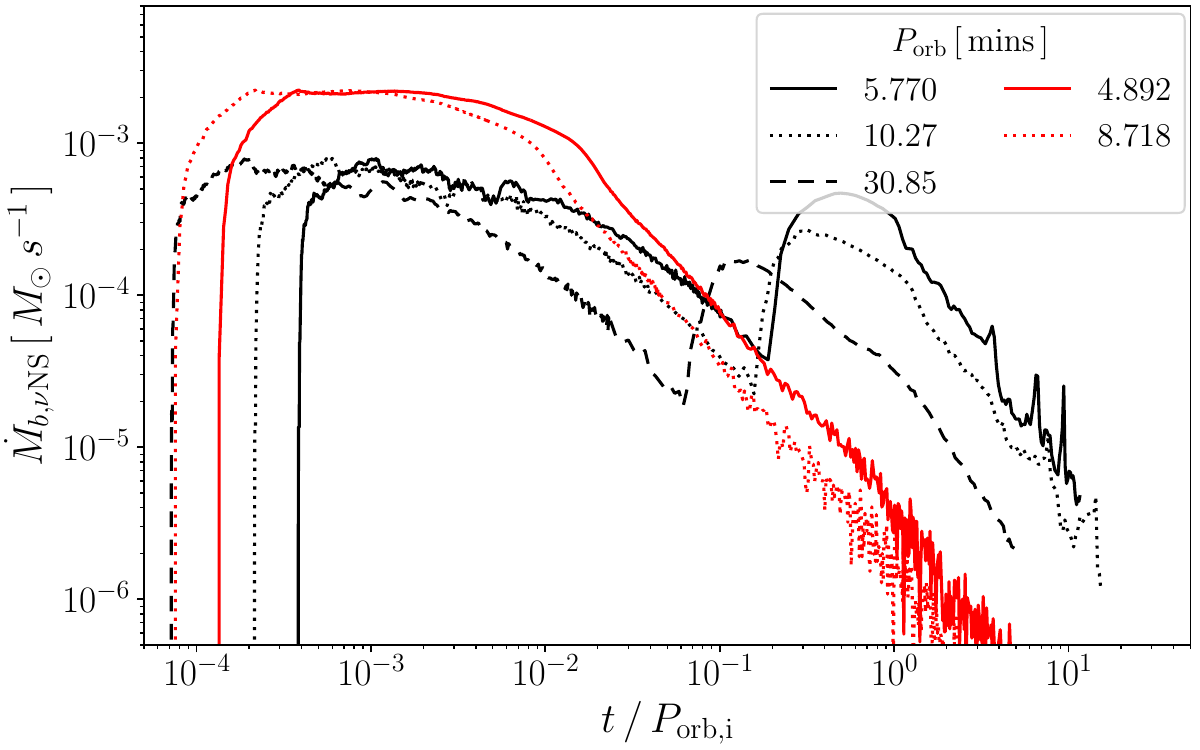}
   \includegraphics[width=0.49\hsize,clip]{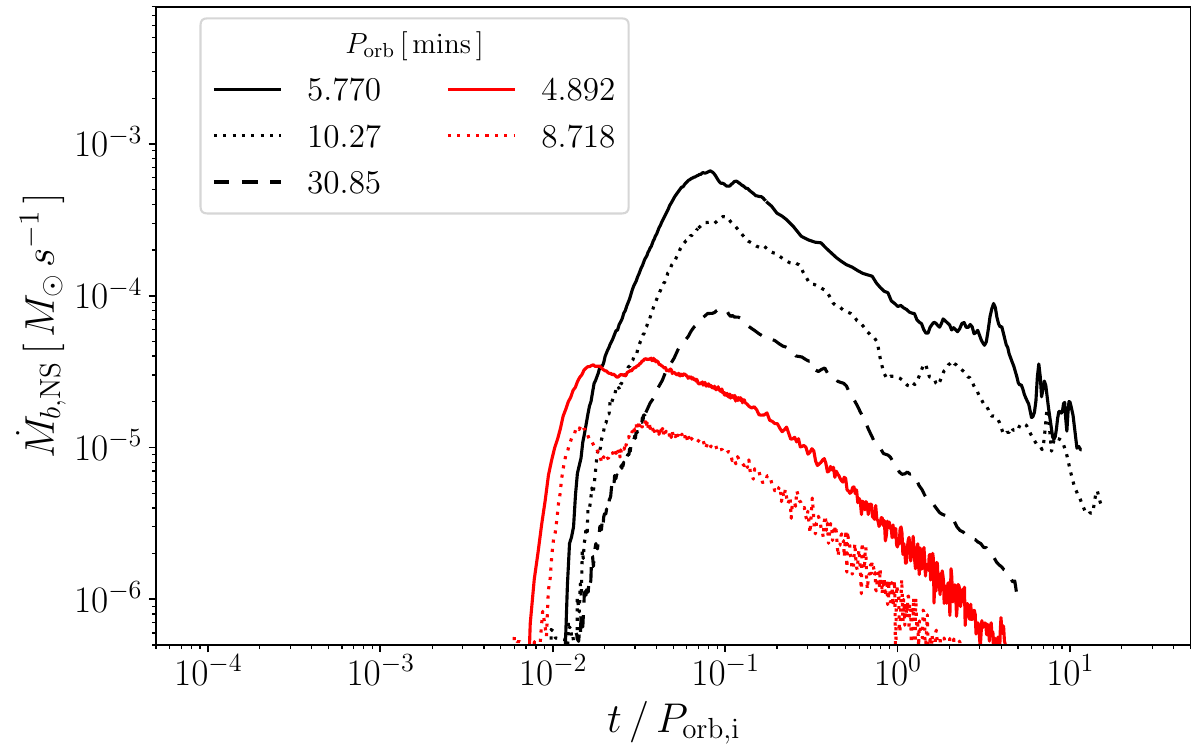}
    \caption{Accretion rate onto the $\nu$NS (left) and the NS companion (right) as a function of time, obtained from SPH simulations of BdHNe with various orbital periods and CO star progenitors. The black curves correspond to binaries formed by a CO star evolved from a ZAMS with $M_{\rm ZAMS}=30 M_\odot$ and a $1.9~M_\odot$ NS companion. {The CO star undergoes collapse, ejecting an SN with an energy of $3.26\times 10^{51}$~ergs}. The red curves correspond to binaries formed by a CO star from a ZAMS with $M_{\rm ZAMS}=15 M_\odot$ and a $1.4~M_\odot$ NS companion,  { where the corresponding SN energy is $1.1\times 10^{51}$~ergs}. We refer to Table \ref{tab:COstar} for the CO-NS binary properties.}
    \label{fig:Mdots_Porb}
\end{figure*}

The top panel of Fig. \ref{fig:3DSPH} shows that the SN ejecta, which is gravitationally captured by the NS companion, first forms a tail behind the star and then circularizes around it, forming a thick disk. At the same time, the particles from the innermost layers of the SN ejecta that could not escape from the $\nu$NS gravitational field fallback are accreted by the $\nu$NS. After a few minutes, part of the material in the disk around the NS companion is also attracted by the $\nu$NS, enhancing the accretion process onto the $\nu$NS. 

The bottom panel of Fig. \ref{fig:3DSPH} corresponds to a simulation with a CO star coming from a $M_{\rm ZAMS}=15~M_\odot$ progenitor with a $1.4~M_\odot$ companion. At the beginning of the simulation, the CO core collapses into a $\nu$NS of $1.4~M_\odot$, and around $1.6~M_\odot$ is ejected by the SN explosion {with a total energy of $1.1\times 10^{51}$~erg.}. Almost all the SN ejecta leave the system without being affected by the NS companion gravitational field.

The hydrodynamics of the matter infalling and accreting onto an NS at hypercritical rates has been extensively studied in different astrophysics contexts taking into account details on the neutrino emission, e.g., fallback accretion in SN \citep{1972SvA....16..209Z,1996ApJ...460..801F,2006ApJ...646L.131F,2009ApJ...699..409F}, accreting NS in X-ray binaries \citep{1973PhRvL..31.1362R}, and for the case of BdHNe, we refer to \citet{2014ApJ...793L..36F,2016ApJ...833..107B,2018ApJ...852..120B} for details. The latter includes a formulation in a general relativistic background and accounts for neutrino flavor oscillations. The relevant, not obvious result is that these simulations show that the NS can accrete the matter at the hypercritical rate at which baryonic mass from the SN ejecta falls into the gravitational capture region of the NS. The simulations show the accretion rate in BdHNe is hypercritical, reaching peak values of up to $10^{-3} M_\odot$ s$^{-1}$. This implies the accretion flow reaches densities $\sim 10^8$ g cm$^{-3}$ near the NS surface (see, e.g., Appendix B in \citealp{2016ApJ...833..107B} and Table 1 in \citealp{2018ApJ...852..120B}), which is about $12$ orders larger than the densities of an Eddington-limited accretion flow in low-mass X-ray binaries.

Therefore, we assume that the accretion rate inferred with the SPH code is the effective baryonic mass accretion rate onto the NS, i.e.,
\begin{equation}
     \dot{M}_{b, \nu \rm NS} = \dot{M}_{\nu \rm NS}^{\rm cap},\quad \dot{M}_{b, \rm NS} =  \dot{M}_{\rm NS}^{\rm cap}.
\end{equation}
Figure \ref{fig:Mdots_Porb} shows the accretion rate onto the $\nu$NS and the NS companion obtained from SPH simulations for selected orbital periods and progenitor for the CO star: $M_{\rm ZAMS}= 30~M_\odot$ and $M_{\rm ZAMS}= 15~M_\odot$, with a NS companion with $1.9~M_\odot$ and $1.4~M_\odot$ respectively (see Table~\ref{tab:COstar}). 

For the systems with the CO progenitor $M_{\rm ZAMS}= 30~M_\odot$, the accretion rate onto the $\nu$NS shows two prominent peaks. The second peak of the fallback accretion onto the $\nu$NS is a unique feature of BdHNe because, as explained above \citep[see, also,][for additional details]{2019ApJ...871...14B}, it is caused by the influence of NS companion. The accretion rate onto the NS companion shows a single-peak structure, accompanied by additional peaks of smaller intensity and shorter timescales. This feature is more evident in short-period binaries. Such small peaks are produced by higher and lower accretion episodes as the NS companion orbits across the ejecta and finds higher and lower-density regions.

For systems with a CO progenitor with $M_{\rm ZAMS}= 15~M_\odot$, the second peak on the mass accretion rate on the $\nu$NS does not occur, while the accretion rate on the NS companion is much lower. This is due to the small mass ejected in the SN explosion and its high velocity.
\begin{table}
    \centering
    \begin{tabular}{c|cccccc}
    \hline
         $M_{\rm ZAMS}$& $M_{\nu \rm NS}$ & $M_{\rm ej} $ & $M_{\rm NS}$ & $P_{\rm orb} $& $\Delta M_{\rm b,NS}$  &$\Delta M_{\rm b,\nu NS}$ \\
         $M_\odot$ & $M_\odot$ & $M_\odot$ &  $M_\odot$ &  ${\rm min} $&$M_\odot$ &  $M_\odot$ \\ \hline\hline
         $30$ & $1.8$ & $7.14$ & $1.9$ & $5.77$& $0.328$ & $0.414$  \\
          &  & &  & $10.27$ & $0.281$ & $0.391$\\
         &  & &  & $30.85$ & $0.117$ & $0.322$\\ \hline
         $15$ & $1.4$ & $1.6$ & $1.4$ & $4.89$ & $0.006$ & $0.029$\\
         &  & &  &$8.72$ &$0.006$  & $0.023$ \\ \hline\hline
    \end{tabular}
    \caption{Properties of the CO-NS binary before the CO core gravitational collapse. The CO star mass is $M_{\rm CO} = M_{\nu \rm NS} + M_{\rm ej}$, where the mass of the CO unstable iron core gives the $\nu$NS mass.}
    \label{tab:COstar}
\end{table}

%%%%%%%%%%%%%%%%%%%%%%%%%%%%%%%%%%%%%%%%%%%%%%%%%%%%%%%%%%%
%%%%%%%%%%%%%%%%%%%%%%%%%%%%%%%%%%%%%%%%%%%%%%%%%%%%%%%%%%%
\section{Equation of state}\label{sec:3}
%%%%%%%%%%%%%%%%%%%%%%%%%%%%%%%%%%%%%%%%%%%%%%%%%%%%%%%%%%%
%%%%%%%%%%%%%%%%%%%%%%%%%%%%%%%%%%%%%%%%%%%%%%%%%%%%%%%%%%%

To consider the scenario proposed in the present paper, nuclear matter equations of state (EOS) that can describe NS macroscopic properties are necessary. In recent decades, many relativistic EOS were proposed and analyzed in light of bulk nuclear matter properties \citep{2014PhRvC..90e5203D}. The detection of massive NSs \citep{2010Natur.467.1081D,2013Sci...340..448A,2020NatAs...4...72C,2021ApJ...915L..12F,2022ApJ...934L..17R}, imposes strong constraints on the density dependence of the EOS and the EOS that passed the test of satisfying nuclear bulk properties were confronted with the observational data in several works, see for instance \citep{2016PhRvC..93b5806D,PhysRevC.94.049901,2016PhRvC..94c5804F}. More recently, the data on GW170817 \citep{2017PhRvL.119p1101A} by the LIGO/Virgo Collaboration have been used to constrain the radius of the canonical NS (i.e., of a mass of 1.4 $M_\odot$) by tidal polarizabilities of the stars involved in the merger \citep[see, e.g.,][]{2019PhRvC..99d5202L}. In the last years, the data from the Neutron star Interior Composition Explorer (NICER) X-ray telescope for a canonical NS \citep{2019ApJ...887L..21R,2019ApJ...887L..24M} and a massive NS \citep{2021ApJ...918L..27R, 2021ApJ...918L..28M} have also been used to constrain the EOS, both measurements imposing further restrictions on the EOS. Another interesting object is the NS in the quiescent low-mass X-ray binary, the NGC 6397 \citep{2001ApJ...563L..53G,2011ApJ...732...88G,2014MNRAS.444..443H}, which provided reliable constraints, as seen in \cite{2016ARA&A..54..401O,2018MNRAS.476..421S}.
It is also worth mentioning other observations that have been used as constraints, but they must be considered carefully due to their specific nature. One refers to a massive, fast black widow with a large error bar \citep{2022ApJ...934L..17R}. {Another one} is the gravitational wave emission resulting from the merger of a BH with another object that can be either the smallest BH or the most massive NS ever detected, GW190814 \citep{2020ApJ...896L..44A}. The last one is a very compact object, perhaps a quark star, known as XMMUJ173203.3--344518 \citep{2022NatAs...6.1444D}. Notice, however, that the present observations are still not very restrictive, as discussed in several works where a Bayesian inference approach has been used to constrain the parameters of relativistic mean-field models  \citep{Traversi:2020aaa,Malik:2022zol,Beznogov:2022rri,Malik:2023mnx,2024MNRAS.529.4650H}.

Having all those considerations in mind, we have chosen to work with two different relativistic mean-field (RMF) parametrizations that satisfy all of the constraints mentioned above, namely eL3$\omega\rho$ \citep{2022CoTPh..74a5302L}, and NL3$\omega\rho$ \citep{Horowitz:2000xj,2016PhRvC..94c5804F,2016PhRvC..94a5808P,2022ApJ...936...41L}. 
One should notice that the Lagrangian density nonlinear terms are presented differently in the literature. The interested reader on a uniform and clear notation can refer to \citet{2023BrJPh..53..137B}. One aspect that remains to be mentioned is the hyperon puzzle. While including hyperons seems to be natural from the theoretical point of view, it softens the EOS with a consequent decrease in the maximum NS mass, possibly not attaining 2$M_\odot$; see, for instance, the discussions in \citep{Malik:2022jqc,Sun:2022yor,Malik:2023mnx} where the inclusion of hyperons has been considered within a Bayesian inference approach. There are different ways to circumvent this problem in the literature, and we cannot say it is completely solved. Hence, one of the models we use next fails to describe the highly massive objects detected so far when hyperons are included. Nevertheless, it remains a good choice if other aspects are considered, as discussed in \cite{2022MNRAS.512.5110L}. {We consider these two EOS as representatives of EOS with similar properties allowing for $2 M_\odot$ stars, also including hyperons}, see for instance \citep{Malik:2023mnx}. 

The Lagrangian density of these models is given by \citep{Horowitz:2000xj,2010PhRvC..82e5803F}:
\begin{align}
\mathcal{L}_{QHD} &= \bar{\psi}_B[\gamma^\mu(i\partial_\mu  - g_{B\omega}\omega_\mu   - g_{B\rho} \frac{1}{2}\vec{\tau} \cdot \vec{\rho}_\mu)  \nonumber \\
&- (M_B - g_{B\sigma}\sigma)]\psi_B 
-U(\sigma)  
  + \frac{1}{2}(\partial_\mu \sigma \partial^\mu \sigma - m_s^2\sigma^2) \nonumber \\
    &- \frac{1}{4}\Omega^{\mu \nu}\Omega_{\mu \nu} + \frac{1}{2} m_v^2 \omega_\mu \omega^\mu
%+\frac{1}{4!}\xi g_v^4 (\omega_\mu \omega^\mu)^2
 - \frac{1}{4}{\bf{P}^{\mu \nu} \cdot \bf{P}_{\mu \nu}}  
 \nonumber \\
 &+ \frac{1}{2} m_\rho^2 \vec{\rho}_\mu \cdot \vec{\rho}^{ \; \mu} \mathcal{L}_{\omega\rho} + \mathcal{L}_{\phi},
\end{align}
in natural units. $\psi_B$ represents the Dirac field, where $B$ can stand either for nucleons only ($N$) or  nucleons ($N$) and hyperons ($H$). The $\sigma$, $\omega_\mu$ and $\vec{\rho}_\mu$ are the mesonic fields and $\vec{\tau}$ are the Pauli matrices. The $g$'s are the Yukawa coupling constants, $M_B$ is the baryon mass, $m_s$, $m_v$  and $m_\rho$ are the masses of the $\sigma$, $\omega$ and $\rho$ mesons respectively. The $U(\sigma)$ is the self-interaction term \citep{1977NuPhA.292..413B}
  \begin{equation}
U(\sigma) =  \frac{1}{3!}\kappa \sigma^3 + \frac{1}{4!}\lambda \sigma^4, 
  \end{equation}
the $\omega^4$ term was introduced in \citep{Sugahara:1993wz}, and it softens the EOS at high densities (see discussion in \citealp{Malik:2023mnx}), and $\mathcal{L}_{\omega\rho}$ is a non-linear $\omega$-$\rho$ coupling interaction discussed in \citep{Horowitz:2000xj}
  \begin{equation}
 \mathcal{L}_{\omega\rho} = \Lambda_{\omega\rho}(g_{N\rho}^2 \vec{\rho^\mu} \cdot \vec{\rho_\mu}) (g_{N\omega}^2 \omega^\mu \omega_\mu) , \label{EL2}
\end{equation}
necessary to correct the slope of the symmetry energy ($L$). The last term $\mathcal{L}_\phi$ is related the hidden strangeness $\phi$ vector meson, which couples only with the hyperons ($H$), not affecting the properties of nuclear matter
\begin{equation}
\mathcal{L}_\phi = g_{H \phi}\bar{\psi}_H(\gamma^\mu\phi_\mu)\psi_H + \frac{1}{2}m_\phi^2\phi_\mu\phi^\mu - \frac{1}{4}\Phi^{\mu\nu}\Phi_{\mu\nu}. \label{EL3} 
\end{equation}
To account for the chemical stability and charge neutrality of NS, leptons have to be considered, and the corresponding Lagrangian density reads
\begin{equation}
 \mathcal{L}_l = \sum_{l=e,\mu}  \bar{\psi}_l[\gamma^\mu(i\partial_\mu  - m_l)]\psi_l, 
\end{equation}
where $l$ represents electrons and muons, and $\beta$-equilibrium conditions establish a relation between the different chemical potentials
\begin{equation}
 \mu_B=\mu_n-q_B \mu_e, \quad \mu_\mu=\mu_e, 
\end{equation}
where $\mu_i$ designates the chemical potential of the baryon ($B$), neutron ($n$), electron ($e$) and muon ($\mu$), and $q_B$ the electric charge of baryon $B$. The imposition of these relations defines the NS composition. In the present study, we consider the newly formed NS, i.e., the $\nu$NS, already in a neutrino-free regime.

Given the Lagrangian density, the equations of motion are obtained and solved with the help of an RMF approximation. All details can be found in the literature \citep[see for instance][]{1992RPPh...55.1855S,Glendenning:1997wn,2022ApJ...936...41L} and are not reproduced here. The first step is to obtain expressions for the pressure ($P_B$), energy density ($\varepsilon_B$), and baryonic density ($n_B$) at zero temperature, which are used as input to the calculation of NS macroscopic properties via Einstein equations. The next step is to calculate the EOS at fixed entropy per baryon ($s_B/n_B$) for neutrino-free matter \citep[e.g.][]{Prakash:1996xs}. For this case, the Fermi distribution functions are no longer step functions, and anti-particles have to be taken into account as
\begin{equation}
    f_{B \pm} = \frac{1}{1+\exp[(E_B \mp \mu^\ast_B)/T]},
\end{equation}
with energy $E_B= \sqrt{k^2 + {M_B^\ast}^2}$, where $M_B^\ast$ is the effective mass and $\mu_B^\ast$ is the
effective chemical potential. The entropy density can be easily obtained from the expression
\begin{equation}
     s_B = \frac{ \varepsilon_B + P_B - 
     \sum_B \mu_B n_B}{T},
\end{equation}
or computed as the entropy density of a free Fermi gas. 

\begin{figure}
    \centering
    \subfigure[]{\includegraphics[width=\hsize,clip]{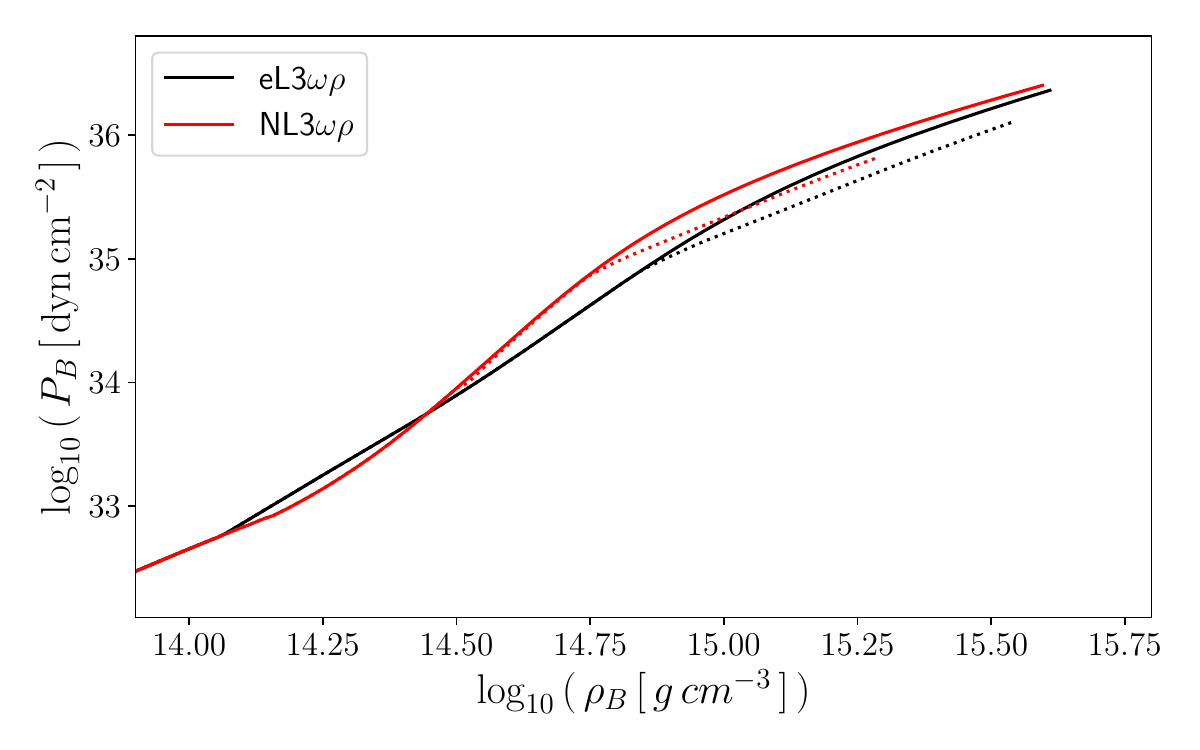}}
   \subfigure[]{\includegraphics[width=\hsize,clip]{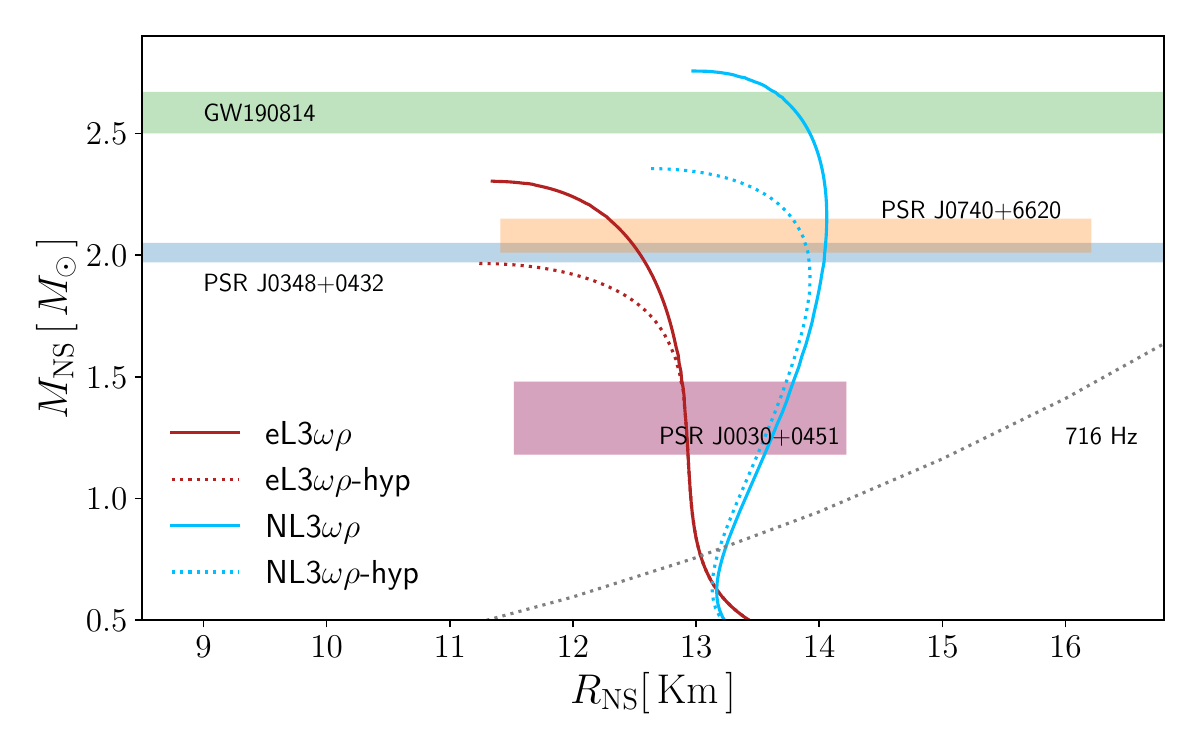}}
   \subfigure[]{\includegraphics[width=\hsize,clip]{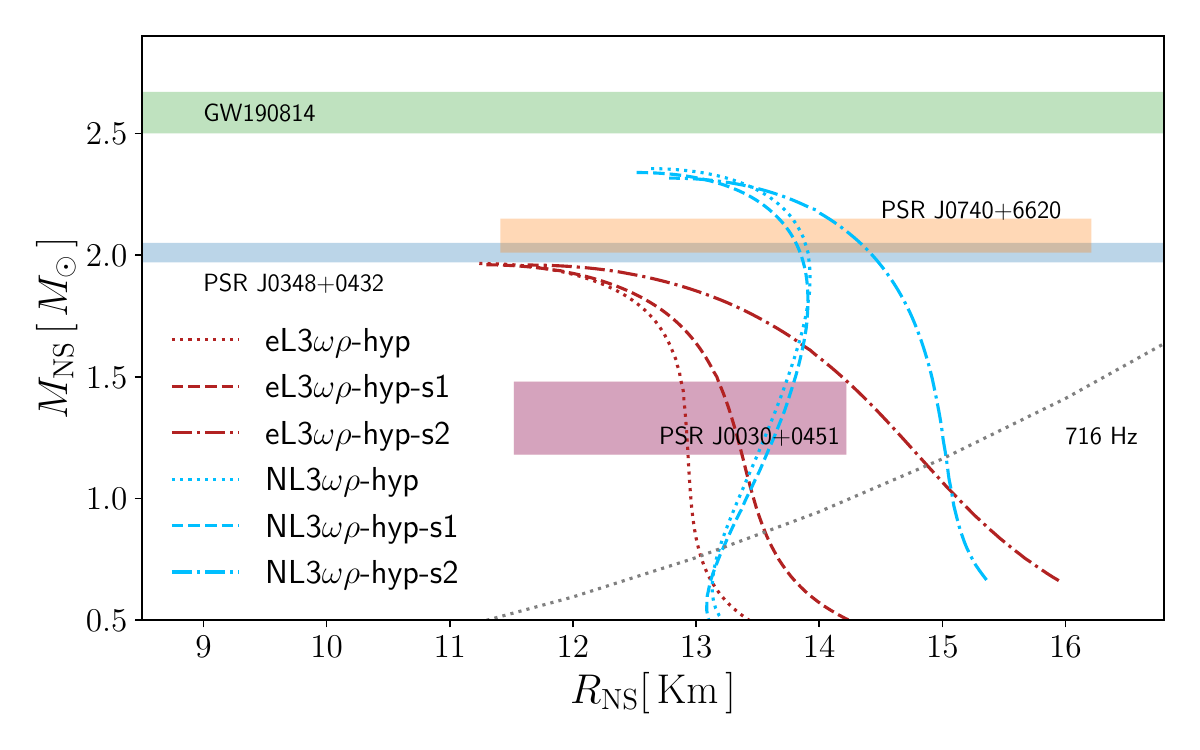}}
    \caption{Top: Pressure-energy density relation for cold NS EOS with the matter with nucleons only (solid line) and nucleons and hyperons (dotted line). Middle: Mass-radius relation for non-rotating NSs. The color bands represent observational constraints given by the pulsar mass of PSR J0348+0432 \citep{2013Sci...340..448A}, on mass and radius from NICER measurements for pulsars PSR J0030+0451 \citep{2019ApJ...887L..21R,2019ApJ...887L..24M} and PSR J0740+662 \citep{2021ApJ...918L..27R,2021ApJ...918L..28M}, and the mass of the secondary compact object of the GW190814 event \citep{2020ApJ...896L..44A}. The dashed gray line corresponds to the fastest observed radio pulsar, PSR J1748--2446ad \citep{2006Sci...311.1901H}. Bottom: Same as middle panel but adding the hot neutron star EOS with nucleons mixed with hyperons for two constant entropy per baryon values, $s_B/n_B = 1, 2$.}
    \label{fig:eos}
\end{figure}
\begin{table}
    \centering
    \begin{tabular}{c|ccc}
    \hline
       EOS  & $M^{j=0}_{\rm max}$ & $M^{\rm Kep}_{\rm max}$ & $\Omega^{\rm Kep}_{\rm max}$  \\
        & ($M_\odot$) & ($M_\odot$) & $10^4$ s$^{-1}$  \\ \hline \hline
        eL3$\omega\rho$ & $2.30$ & $2.75$ & $1.08$\\
        eL3$\omega\rho$-hyperons & $1.96$ & $2.35$ & $0.98$\\
        eL3$\omega\rho$-hyperons-S1 & $1.96$ & $2.32$ & $0.97$\\
        eL3$\omega\rho$-hyperons-S2 & $1.96$ & $2.27$ & $0.92$\\
          NL3$\omega\rho$ & $2.76$ &$3.36$& $0.98$ \\
          NL3$\omega\rho$-hyperons &$2.35$  & $2.88$ &$0.91$ \\
         NL3$\omega\rho$-hyperons-S1 &$2.34$  & $2.84$ &$0.91$ \\
          NL3$\omega\rho$-hyperons-S2&$2.32$  & $2.75$ &$0.88$ \\
            \hline \hline
    \end{tabular}
    \caption{For selected EOS, the table lists the maximum stable mass for the non-rotating and uniformly rotating configurations, $M_{\rm max}^{j=0}$ and $M_{\rm max}^{\rm Kep}$, and the maximum rotation frequency, $\Omega_{\rm max}^{\rm Kep}$. S1 and S2 stand for the two constant values of entropy per baryon, $s_B/n_B = 1$, $2$.}
    \label{tab:EoS_mass}
\end{table}

The top panel of Fig.~\ref{fig:eos}  shows the pressure as a function of the energy density for the eL3$\omega\rho$ and NL3$\omega\rho$ EOS at zero temperature. For both parameterizations, we considered the case of matter formed only by nucleons and the case of nucleons mixed with hyperons. The middle panel of Fig.~\ref{fig:eos} shows the mass-radius relation for the non-rotating and cold neutron star configurations (see also Table~\ref{tab:EoS_mass}), obtained including a crust. We have considered the BPS model for the outer crust \citep{Baym:1971pw} and a Thomas-Fermi calculation of the inner crust \citep{2008PhRvC..78a5802A,2012PhRvC..85e5808G}. The core-crust transition occurs at $n_B=0.082$ fm$^{-3}$ ($\rho_B = \varepsilon_B/c^2=1.39\times 10^{14}$~g/cm$^3$) for the NL3$\omega\rho$ model. As discussed in \citet{2016PhRvC..94c5804F}, it is important to consider an inner crust EoS described by the same model as the core. We will take for both EOS the same crust model because they have similar symmetry energy at sub-saturation densities, and, therefore, it is expected that the two inner crust EoS do not differ much \citep{2016PhRvC..94a5808P}. The bottom panel of Fig.~\ref{fig:eos} shows the mass-radius relation for non-rotating and hot neutron star configurations for the  eL3$\omega\rho$ and NL3$\omega\rho$ EOS parametrization with nucleons mixed with hyperons. The hot EOS generally produces less compact configurations than the cold one, but the maximum allowed mass remains roughly the same for both cases (see also Table~\ref{tab:EoS_mass}). In the last two plots, we have also colored the regions corresponding to the observational constraints discussed at the beginning of this section. We have used two constant values of entropy per baryon, $s_B/n_B = 1, 2$, which we designate as S1 and S2 in Table~\ref{tab:EoS_mass} and Fig.~\ref{fig:eos}.

%%%%%%%%%%%%%%%%%%%%%%%%%%%%%%%%%%%%%%%%%%%%%%%%%%%%%%%%%%%
%%%%%%%%%%%%%%%%%%%%%%%%%%%%%%%%%%%%%%%%%%%%%%%%%%%%%%%%%%%
\section{Evolution of the neutron star structure}\label{sec:4}
%%%%%%%%%%%%%%%%%%%%%%%%%%%%%%%%%%%%%%%%%%%%%%%%%%%%%%%%%%%
%%%%%%%%%%%%%%%%%%%%%%%%%%%%%%%%%%%%%%%%%%%%%%%%%%%%%%%%%%%

To calculate the time evolution of the $\nu$NS and the NS companion structure during the accretion process, we have implemented a code that uses the {\sc RNS code} \citep{1995ApJ...444..306S} (with the quadrupole correction performed in \citealp{2015PhRvD..92b3007C}). Given the EOS, the code calculates the stable, rigidly rotating, corresponding NS configuration of equilibrium in axial symmetry for the baryonic mass, $M_b$, and angular momentum, $J$, at a given time. The value of $M_b$ is updated using the baryonic mass accretion rate from the SPH numerical simulation (see Section \ref{sec:2}). The value of $J$ is obtained by angular momentum conservation \citep{2022PhRvD.106h3002B}:
\begin{equation}\label{eq:Jdot}
\dot{J}= \tau_{\rm acc} +  \tau_{\rm mag},
\end{equation}
where the torques acting onto the stars are specified as follows. Our numerical simulations indicate that the infalling material forms a disk around the star before being accreted. Therefore, the accreted matter exerts a (positive) torque onto the star
\begin{equation}\label{eq:chi}
  \tau_{\rm acc} = \chi\,l\,\dot{M}_b,
\end{equation}
where $\chi \leq 1$ is an efficiency parameter of angular momentum transfer, and $l$ is the specific (i.e., per unit mass) angular momentum of the inner disk radius, $R_{\rm in}$, and it is given by
\begin{equation}
    l = \begin{cases}
    l_{\rm isco} ,& \text{if } R_{\rm in}\geq R_{\rm NS}, \\
    \Omega R_{\rm NS}^2,              & \text{if } R_{\rm in}< R_{\rm NS},
\end{cases}
\end{equation}
with $l_{\rm isco}$ the specific angular momentum of the innermost stable circular orbit around the NS, while $R_{\rm NS}$ and $\Omega$ are, respectively, the NS radius and angular velocity. All these quantities are obtained from the numerical solution of the Einstein equations directly from the {\sc RNS code}. It is worth noting that due to the high accretion rates (see Fig.~\ref{fig:Mdots_Porb}), the NSs evolve in the regime where $R_{\rm mag}< R_{\rm NS}$, with $R_{\rm mag}$, the magnetospheric radius \citep{1972A&A....21....1P}:
\begin{equation}
    R_{\rm mag}=\left(\frac{\mu_{\rm dip}^2}{\dot{M}\sqrt{2GM_{\rm NS}}}\right)^{2/7},
\end{equation}
where $\mu_{\rm dip}= B_{\rm dip} R_{\rm NS}^3$ is the dipole magnetic moment. Additionally, if the magnetic field is not buried by the accretion \citep[e.g.][]{2007MNRAS.376..609P}, the star is subjected to the (negative) torque by the magnetic field. We adopt the dipole+quadrupole magnetic field model \cite[see][for details]{2015MNRAS.450..714P}
\begin{equation}\label{eq:Lsd}
    \tau_{\rm mag} = -\frac{2}{3} \frac{\mu_{\rm dip}^2 \Omega^3}{c^3} \sin^2\theta_1 \left( 1 + \eta^2 \frac{16}{45} \frac{R_{\rm NS}^2 \Omega^2}{c^2} \right),
\end{equation}
where $\eta$ defines the quadrupole-to-dipole magnetic field strength ratio
\begin{equation}\label{eq:eta}
    \eta \equiv \sqrt{\cos^2\theta_2+10\sin^2\theta_2} \frac{B_{\rm quad}}{B_{\rm dip}}.
\end{equation}
In this model, the $m = 0$ mode is set by $\theta_1 = 0$ and any value of $\theta_2$, the $m = 1$ mode is given by $(\theta_1, \theta_2) = (90^\circ, 0^\circ)$, and the $m = 2$ mode by $(\theta_1, \theta_2) = (90^\circ, 90^\circ)$.

In the following, we analyze systems with a CO star evolving from a progenitor with $M_{\rm ZAMS}=30~M_\odot$. For the case of the CO star with a progenitor with $M_{\rm ZAMS}=15~M_\odot$, we do not find any initial condition in which the star evolves outside the stability zone. These systems remain gravitationally bound after the SN explosion and produce binary neutron star systems that will eventually merge, driven by gravitational-wave emission. 
\begin{figure*}
    \includegraphics[width=\hsize,clip]{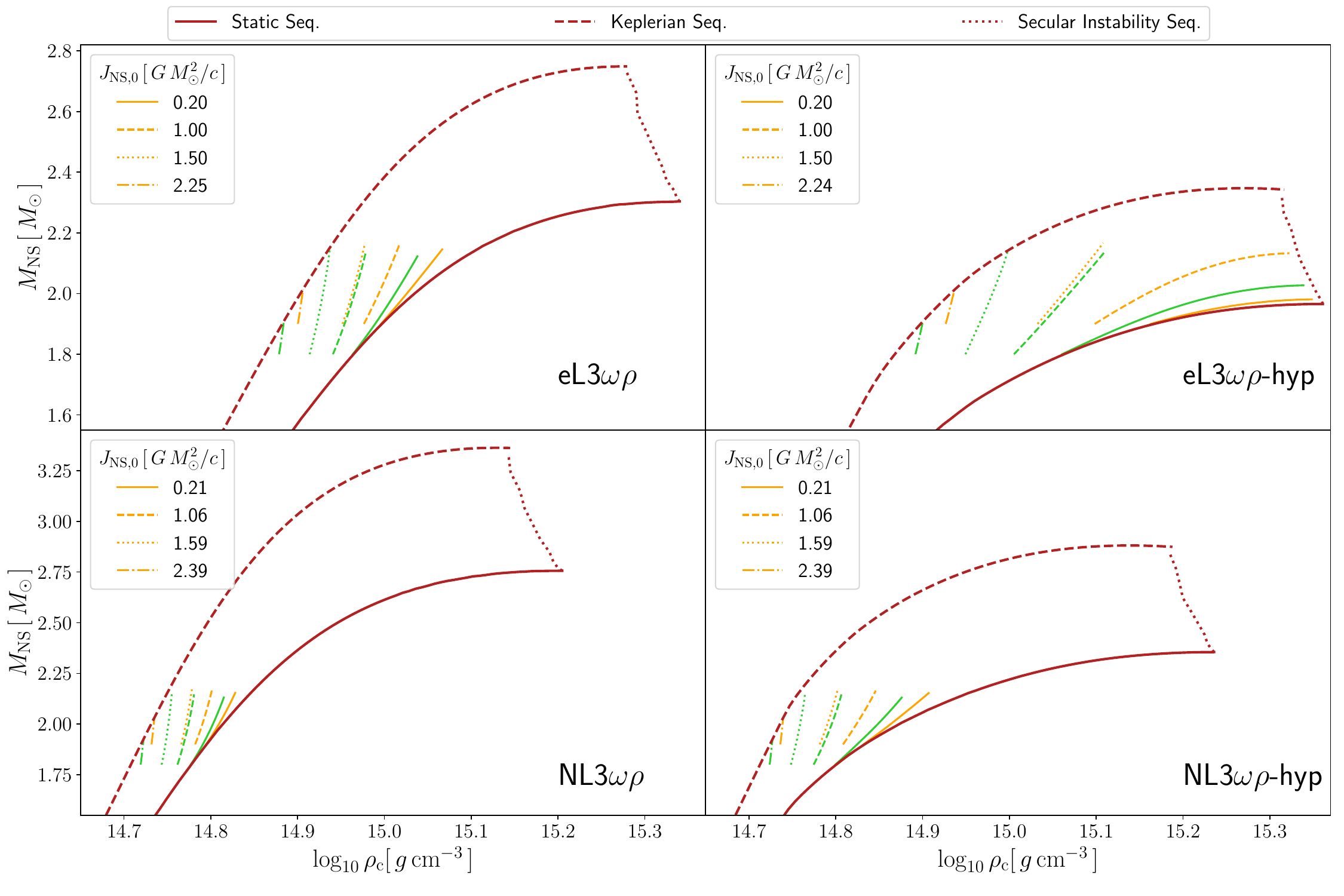}
    \caption{Evolution of the NS ({orange} lines) and the $\nu$NS ({green} lines) in the $\rho_c$-$M_{\rm NS}$ plane for different initial values of  angular momentum, $J$. Each panel corresponds to each EOS summarized in Table~\ref{tab:EoS_mass}. We set the strength of the magnetic field to $B_{\rm ns}=10^{13}$~G and {angular momentum transfer efficiency} $\chi=0.5$. The initial binary system is formed by a $1.9~M_\odot$ NS and a CO, whose progenitor is a star with $M_{\rm ZAMS}=30~M_\odot$, {and the orbital period is $P_{\rm orb} = 5.77$ min}. The stability zone is delimited by the static (red-solid line), Keplerian (red-dashed line), and secular instability sequences (red-dotted line).
}\label{fig:Mrhoevolution}
\end{figure*}

%%%%%%%%%%%%%%%%%%%%%%%%%%%%%%%%%%%%%%%%%%%%%%%%%%%%%%%%%%%
%%%%%%%%%%%%%%%%%%%%%%%%%%%%%%%%%%%%%%%%%%%%%%%%%%%%%%%%%%%
\section{Results}\label{sec:5}
%%%%%%%%%%%%%%%%%%%%%%%%%%%%%%%%%%%%%%%%%%%%%%%%%%%%%%%%%%%
%%%%%%%%%%%%%%%%%%%%%%%%%%%%%%%%%%%%%%%%%%%%%%%%%%%%%%%%%%%

%%%%%%%%%%%%%%%%%%%%%%%%%%%%%%%%%%%%%%%%%%%%%%%%%%%%%%%%%%%
\subsection{$\nu$NS and NS evolution}\label{sec:5a}
%%%%%%%%%%%%%%%%%%%%%%%%%%%%%%%%%%%%%%%%%%%%%%%%%%%%%%%%%%%

Figure~\ref{fig:Mrhoevolution} shows, for the EOS summarized in Table~\ref{tab:EoS_mass}, the evolution of the NS and the $\nu$NS in the central density, $\rho_c$ -- gravitational mass, $M_{\rm NS}$ plane, where $\rho_c$ is $\rho_B = \varepsilon_B/c^2$ at the stellar center, $r=0$. The baryonic mass accretion rate on the NS is taken from the SPH simulation of the BdHN event, which begins with the  SN explosion of a CO star in a binary system with a 1.9~$M_\odot$ NS companion. The CO star evolved from a $M_{\rm ZAMS}=30M_\odot$. When the CO core collapses, it leaves a $1.85 M_\odot$ proto-NS and ejects about $7~M_\odot$ in the SN explosion. {The orbital period is $P_{\rm orb} = 5.77$ min.} The star increases its central density in all cases as it accretes baryonic mass and angular momentum. The equatorial radius and angular velocity also increase when the secular instability limit is not reached.

For sufficiently high initial angular momentum, the stars reach the mass-shedding limit. As shown by 3D numerical simulations of uniformly rotating NSs by \citet{2000PhRvD..61d4012S}, mass-shedding limit leads to a dynamical instability point near secular instability, which, in turn, leads to gravitational collapse into a BH. They show this situation will occur by an NS accreting at the mass-shedding limit. Therefore, we here assume the mass-shedding limit and the secular instability as points of BH formation.  {Configurations that can reach this limit are discussed in the next section.}
\begin{figure*}
    \includegraphics[width=\hsize,clip]{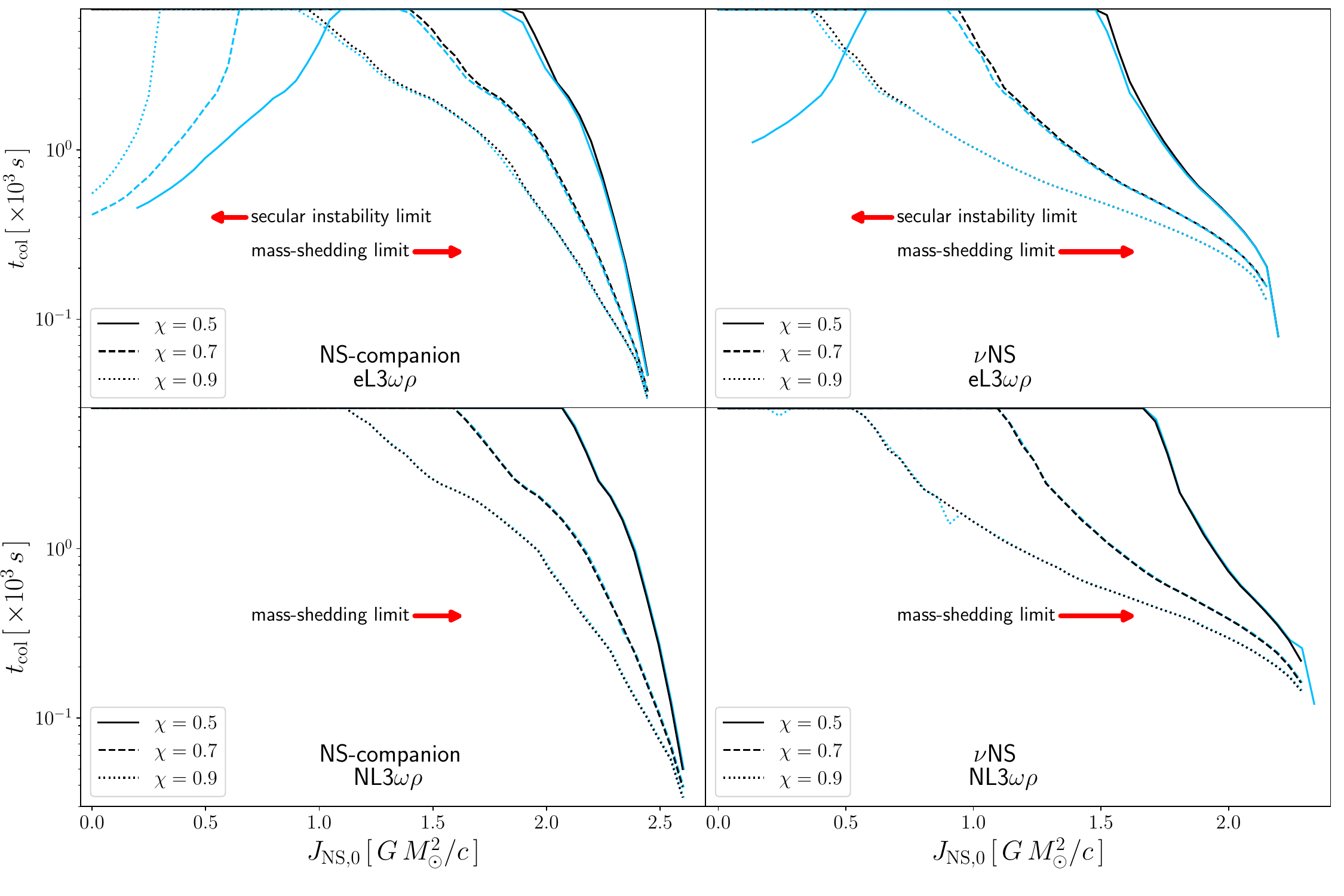} 
    \caption{Time to collapse, defined as the time it takes the stars to leave the stability zone, as a function of the star's initial angular momentum, for different values of the angular momentum efficiency, $\chi$. The upper panels show the results assuming the eL3$\omega\rho$ parametrization for the EoS, while for the lower panels, we use the NL3$\omega\rho$ one. Black lines correspond to matter with nucleons, and blue lines correspond to matter with nucleons mixed with hyperons.The initial binary is the same as in Fig.~\ref{fig:Mrhoevolution}.}\label{fig:CollapseTime}
\end{figure*}
%

%%%%%%%%%%%%%%%%%%%%%%%%%%%%%%%%%%%%%%%%%%%%%%%%%%%%%%%%%%%
\subsection{Time to BH formation}\label{sec:5b}
%%%%%%%%%%%%%%%%%%%%%%%%%%%%%%%%%%%%%%%%%%%%%%%%%%%%%%%%%%%

We now analyze how the initial star's angular momentum affects the occurrence of gravitational collapse by accretion. For the same binary systems of Fig.~\ref{fig:Mrhoevolution}, Fig. \ref{fig:CollapseTime} shows the time taken for the stars to become unstable during the accretion process and probably collapse to a BH as a function of the star's initial angular momentum, for different values for the {angular momentum transfer efficiency, $\chi$. We recall the orbital period is $P_{\rm orb} = 5.77$ min.} The NS companion only reaches the secular instability limit when the eL3$\omega\rho$-EOS with hyperons is assumed. The time the star needs to reach it shortens for larger initial angular momentum or larger $\chi$. On the other hand, independently of the EOS assumed, for $j_{\rm NS,0}\equiv cJ/(GM_\odot^2)\gtrsim 1$ and $\chi=0.9$, the star reaches the mass-shedding, and the time it takes to do it decreases when its initial angular momentum increases. Under certain conditions, with high initial angular momentum, the star could reach the mass shedding limit in less than $50$~s. In contrast, an initially non-rotating star needs more than $300$~s to reach the secular instability limit. The $\nu$NS reaches the secular instability limit for the eL3$\omega\rho$-EOS with hyperons and {only for } $\chi=0.5$. For the $\nu$NS, the mass-shedding limit can be reached for $j_{\rm NS,0}\gtrsim 0.5$, if $\chi=0.9$, {for example}.%{($\chi \lesssim 0.9$ ?)}.

\begin{figure}[t!]
    \includegraphics[width=\hsize,clip]{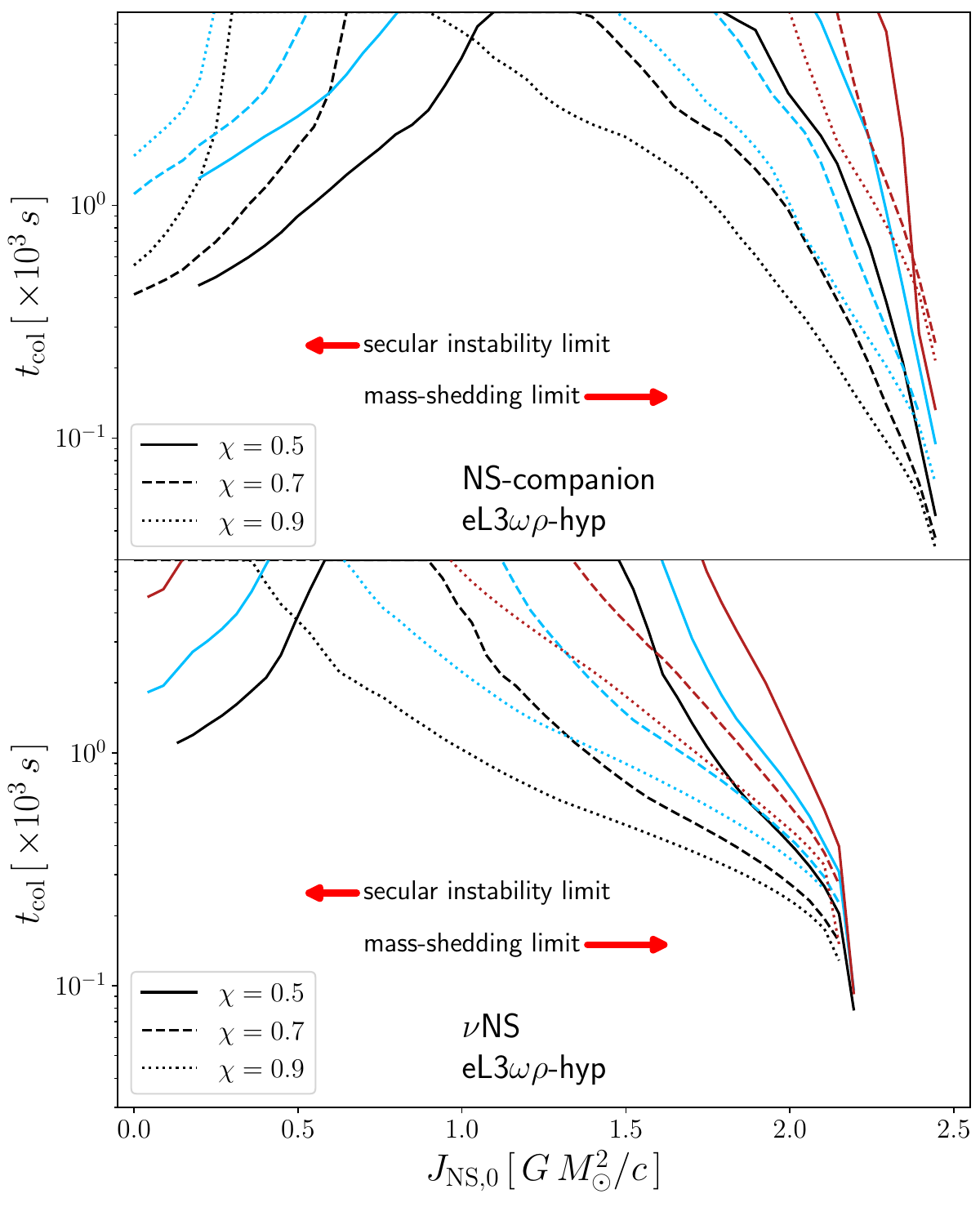} 
    \caption{Same as Fig.~\ref{fig:CollapseTime} but using a EOS with the eL3$\omega\rho$ parametrization with hyperons. The initial binary system is formed by a $1.9~M_\odot$ NS and a CO, whose progenitor is a star with $M_{\rm ZAMS}=30~M_\odot$. The initial binary period is about $5.7$~min for the black lines, $10.2$~min for the blue ones, and $31$~min for the red ones.}\label{fig:CollapseTime_Period}
\end{figure}

Fig.~\ref{fig:CollapseTime_Period} shows the collapse time as done in Fig.~\ref{fig:CollapseTime}, for the same binary system stars but with different initial binary periods (see Fig~\ref{fig:Mdots_Porb}) and employing the  eL3$\omega\rho$-EOS parameterization with nucleons mixed with hyperons.  As shown in Fig.~\ref{fig:Mdots_Porb}, an increase in the initial binary period results in a decrease in the accretion rate of the binary stars, which subsequently requires more time to reach an instability limit. When the initial binary period is greater than $31$~min, the NS companion does not reach the secular instability limit for any value of the $\chi$-parameter, while the $\nu$NS reaches it only for $\chi<0.5$. The initial angular momentum required for the stars to reach the mass-shedding limit increases with the initial binary period. It is worth noting that the {mass loss by the} SN explosion disrupts the final binary system only for the larger initial binary period ($P_{\rm orb}>31$~min). In contrast, the system remains gravitationally bound for the two shorter ones, forming BH-BH or NS-NS binaries. {We refer to \citealp{2024arXiv240115702B} for the latest simulations of the unbinding process in BdHN systems and section \ref{sec:6b} for further discussions on the implications.}

\begin{figure}[t!]
     \includegraphics[width=\hsize,clip]{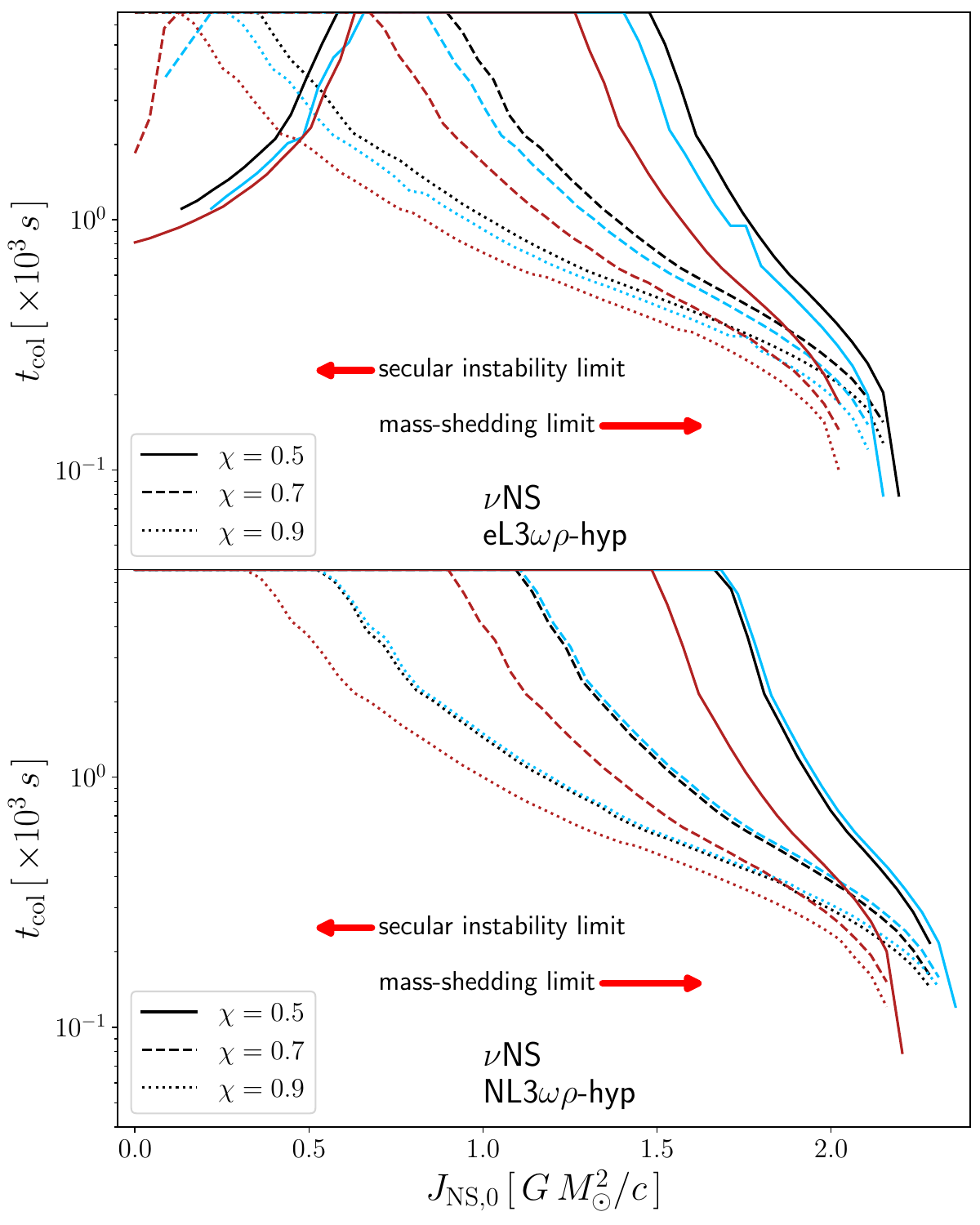} 
    \caption{Same as Fig.~\ref{fig:CollapseTime} but for a hot neutron star using  the eL3$\omega\rho$ and NL3$\omega\rho$ parameterization with hyperons for the EOS. The black lines correspond to the cold EOS, the blue ones for an EOS with constant entropy $S1 (s_B /n_B = 1)$, and the red ones for an EOS with constant entropy $S2(s_B /n_B = 2)>S1$. The initial binary system is formed by a $1.9~M_\odot$ NS and a CO, whose progenitor is a star with $M_{\rm ZAMS}=30~M_\odot$ in an initial binary period of about $5.7$~min.}\label{fig:CollapseTime_entropy}
\end{figure}

Fig.~\ref{fig:CollapseTime_entropy} shows the collapse time for the $\nu$NS described by a hot EOS with uniform entropy (see also bottom panel of Fig.~\ref{fig:eos}). For these cases, the hotter the star, the mass-shedding limit or the secular instability limit is reached in less time, independent of the EOS parameterization used.  For the eL3$\omega\rho$ parameterization, hotter stars can get the secular instability limit with greater values for the {angular momentum transfer efficiency, $\chi$,} when cold stars do not. 

%%%%%%%%%%%%%%%%%%%%%%%%%%%%%%%%%%%%%%%%%%%%%%%%%%%%%%%%%%%
\subsection{Delayed BH formation}\label{sec:5c}
%%%%%%%%%%%%%%%%%%%%%%%%%%%%%%%%%%%%%%%%%%%%%%%%%%%%%%%%%%%

When the stars do not collapse and if the magnetic field is not buried by the accretion~\citep[e.g.][]{2007MNRAS.376..609P}, they will continue losing angular momentum driven by the magnetic field torque after the end of the accretion process. When the final mass of the stars is smaller than the maximum mass allowed for a static star, $M^{\rm max}_{\rm TOV}$, they evolve towards the static sequences as they lose angular momentum. But, when the final mass of either star is larger than $M^{\rm max}_{\rm TOV}$, the stars evolve toward the secular instability limit, leading to a long-delayed collapse into a BH. This will be the case for the stars that do not collapse in the accretion process. On the other hand, if the binary remains gravitationally bound, gravitational-wave emission makes the stars merge in a time $t_{\rm GW}$ (see below). Thus, for the BdHN event to occur, the collapse delay time must be shorter than the merger time. We refer to section \ref{sec:6a} for further discussion on the consequences of the delayed collapse for the BdHN classification. 

To assess the above scenario, we analyze as specific examples the stars described by the eL3$\omega\rho$ EOS with hyperons (Fig.~\ref{fig:Mrhoevolution} and top panel of Fig. \ref{fig:CollapseTime}). 

\begin{figure}
    \centering
    \includegraphics[width=\hsize,clip]{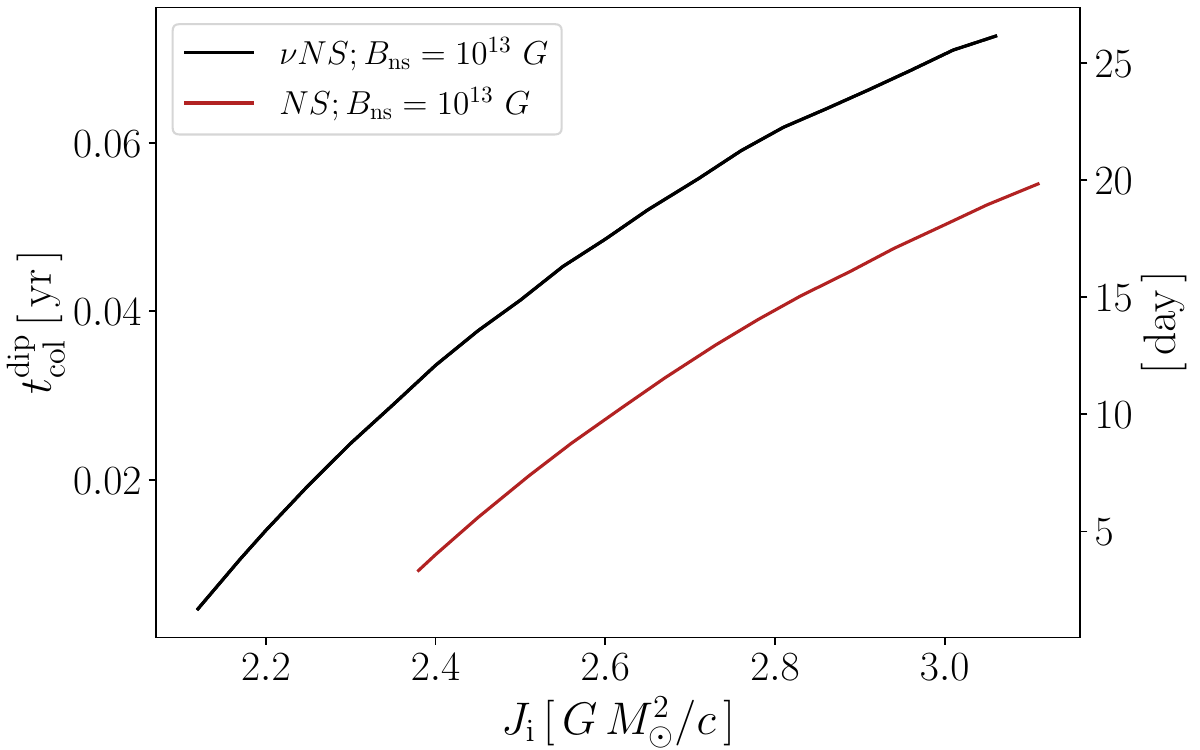}
    \caption{Time to the secular instability limit by magnetic-dipole braking as a function of the initial angular momentum of the $\nu$NS and the NS companion, which do not collapse to a BH in the BdHN event but reach a mass greater than $M_{\rm TOV}^{\rm max}$. The EOS is the eL3$\omega\rho$ parameterization with hyperons. The star's initial mass is about $2.1~M_\odot$, and its radius is between $13$ and $16$~km, depending on its angular momentum. The star reaches secular instability when $J = J_{\rm sec}\approx 3 GM_\odot^2/c$.}
    \label{fig:tcol_dip}
\end{figure}

The time to reach the secular instability limit by magnetic-dipole braking is
\begin{equation}\label{eq:tcolB}
    t^{\rm dip}_{\rm col} = -\frac{3}{2} \frac{c^3}{B^2 } \int_{J_i}^{J_{\rm sec}}\frac{dJ}{R^6\Omega^3},
\end{equation}
where $J_i$ and $J_{\rm sec}$ are the NS angular momentum at the end of accretion and the secularly unstable configuration with the same baryon mass. Equation (\ref{eq:tcolB}) refers to a spherical dipole, so we approximate its radius with the authalic radius, $R\approx (2 R_{\rm eq}+R_p)/3$. The merger time is \citep[see, e.g., Eq. (4.135) in][]{maggiore2007gravitational}
\begin{equation}\label{eq:tauGW}
    t_{\rm GW} = \frac{c^5}{G^3}\frac{5}{256} \frac{a_{\rm orb}^4}{\mu M^2} \frac{48}{19}\frac{1}{g(e)^4}\int_0^e \frac{g(e)^4 (1-e^2)^{5/2}}{e (1 + \frac{121}{304}e^2)}de,
\end{equation}
where $g(e) = e^{12/19}(1-e^2)^{-1}(1+121 e^2/304)^{870/2299}$, $a_{\rm orb}$ is the orbital separation at the end of the accretion process, $M$ and $\mu$ are total and reduced binary mass. The BdHN event of an initial binary with 1.9$M_\odot$ NS and a CO star from a $M_{\rm ZAMS}=30 M_\odot$ progenitor, leaves a binary of approximately equal masses ($\sim 2.1M_\odot$), orbital separation $a_{\rm orb} \sim 3.3 \times 10^{10}$~cm, and eccentricity $e = 0.71$. Equation (\ref{eq:tauGW}) says this system will merge in $t_{\rm GW} = 33$~kyr. Figure~\ref{fig:tcol_dip} shows that $t^{\rm dip}_{\rm col}$ is of the order of $0.05$~yr ($\approx 18$ days) for a magnetic field strength of $10^{13}$~G. Equation ~(\ref{eq:tcolB}) shows this time increases by decreasing the square of the magnetic field strength. For the above numbers, we obtain these stars will gravitationally collapse to a BH before the binary merges for magnetic fields above $10^{10}$~G. Given the shortness of $t^{\rm dip}_{\rm col}$ relative to $t_{\rm GW}$, we can conclude that in $\sim 33$ kyr, this system leads to a BH-BH merger.

%

%%%%%%%%%%%%%%%%%%%%%%%%%%%%%%%%%%%%%%%%%%%%%%%%%%%%%%%%%%
%%%%%%%%%%%%%%%%%%%%%%%%%%%%%%%%%%%%%%%%%%%%%%%%%%%%%%%%%%
\section{Discussion}\label{sec:6}
%%%%%%%%%%%%%%%%%%%%%%%%%%%%%%%%%%%%%%%%%%%%%%%%%%%%%%%%%%
%%%%%%%%%%%%%%%%%%%%%%%%%%%%%%%%%%%%%%%%%%%%%%%%%%%%%%%%%%

%%%%%%%%%%%%%%%%%%%%%%%%%%%%%%%%%%%%%%%%%%%%%%%%%%%%%%%%%%%
\subsection{CO-NS parameters and BdHN types}\label{sec:6a}
%%%%%%%%%%%%%%%%%%%%%%%%%%%%%%%%%%%%%%%%%%%%%%%%%%%%%%%%%%%

We recalled in Section \ref{sec:1} that the various possible fates of the binary imply the three types of sources, BdHNe I, II, and III. In particular, the main parameters of the classification {are the GRB energy and the orbital period of the CO-NS progenitor. These two parameters are naturally} connected in the BdHN model \citep{2016ApJ...832..136R,2016ApJ...833..107B,2019ApJ...871...14B}, as follows.

The minimum energy release of a BdHN is the energy released by the NS companion accretion process when increasing its mass from $M_{\rm NS,0}$ to $M_{\rm NS,f}$, i.e., $\Delta E_{\rm acc} \approx \eta_{\rm acc} \Delta M_{\rm acc} c^2$, where $\Delta M_{\rm acc} = M_{\rm NS,f} - M_{\rm NS,0}$, and $\eta_{\rm acc}$ is the efficiency in converting gravitational into electromagnetic energy. Because the BH forms when the NS reaches the critical mass for gravitational collapse, $M_{\rm crit}$, the energy released to the BH formation trigger point, assuming typical order-of-magnitude values $\Delta M_{\rm acc} = M_{\rm crit} - M_{\rm NS,0} \sim 0.1 M_\odot$ and $\eta_{\rm acc} \sim 0.1$, one obtains $\Delta E_{\rm acc} \sim 10^{52}$ erg. This estimate shows the reason the BdHN I are expected to explain the most energetic GRBs with $E_{\rm iso} \gtrsim 10^{52}$ erg. The connection with the CO-NS progenitor orbital period follows from the relation between the latter and the accreted mass. \citet{2019ApJ...871...14B} showed that for a given SN explosion energy, the amount of accreted matter approximately scales as $\Delta M_{\rm acc} \propto M_{\rm NS,0}^2/P_{\rm orb}^{1/3}$. This expression implies there is a maximum orbital period, $P_{\rm orb,max} \propto M_{\rm NS,0}^6/(M_{\rm crit}-M_{\rm NS,0})^3$, over which no BH is formed by the accretion process. It turns out that $P_{\rm orb,max}$ is of the order of a few minutes for typical CO-NS parameters. The above suggests that the shorter the orbital period, the shorter the time the NS takes to reach the collapse point. The peak time of accretion scales as $t_{\rm peak} \propto P_{\rm orb}^{2/3}$ \citep{2019ApJ...871...14B}. Besides, there is a strong dependence on the initial NS angular momentum, as shown in this article (e.g., Figs. \ref{fig:CollapseTime} and \ref{fig:CollapseTime_Period}). Examples of BdHNe I are GRB 130427A \citep{2019ApJ...886...82R}, GRB~180720B \citep{2022ApJ...939...62R}, and GRB 190114C \citep{2021PhRvD.104f3043M, 2021A&A...649A..75M}. The BdHNe I show seven observable emission episodes in the sequence of physical processes triggered by the SN explosion in the CO-NS: GRB precursors, MeV prompt, GeV and TeV emissions, X-optical-radio afterglow, and optical SN emission. These emissions involve the physics of the early SN, NS accretion, BH formation, synchro-curvature radiation, and quantum and classic electrodynamics processes. We refer to \citet{2023ApJ...955...93A} and the Appendix in \citet{2024ApJ...966..219B} for details.

Therefore, binaries with $P_{\rm orb} > P_{\rm orb,max}$ do not form a BH. These binaries lead to the subclasses BdHNe II and III. The separatrix between these two subclasses is given by the fact that as the period and orbital separation increase, the role of the NS companion diminishes, becoming negligible for binaries with periods of hours. Thus, BdHNe III are expected to release an energy similar to that of a single core-collapse event without any companion. Accretion is expected only from fallback onto the $\nu$NS, which has a maximum peak accretion $10^{-3} M_\odot$ s$^{-1}$ for about seconds (e.g., Fig. \ref{fig:Mdots_Porb}), so $\Delta M_{\rm acc}\sim 10^{-3} M_\odot$, leading to $\Delta E_{\rm acc} \sim 10^{50}$ erg. We refer to \citet{2023ApJ...945...95W} for the detailed analysis of GRB~171205A as an example of BdHN III.

The sources with energies in the range $10^{50}$--$10^{52}$ erg are explained by the BdHNe II, with orbital periods from tens of minutes to hours. We refer to \citet{2022ApJ...936..190W} for the analysis of GRB~190829A as a BdHN II. At this stage, it is worth recalling that section \ref{sec:5c} has shown that the NS companion, in some cases, could lie at the supramassive metastable region at the end of the accretion phase. For those systems, delayed BH formation is expected to occur under the action of braking mechanisms, e.g., by a magnetic field. A natural question arises as to whether these systems should be classified as BdHN I or II. Figure \ref{fig:tcol_dip} shows that the delay time of BH formation could be a few $10^6$ s ({$10$ days} or longer for magnetic fields lower than $10^{13}$ G) after the SN breakout. Whether the BH formation implies BdHN I signatures in a binary that would lead to a BdHN II depends on the emission processes related to the BH and the differences in the system properties relative to a BdHN I. In this sense, the delay time could be crucial (e.g., the density surrounding the system decreases with time). It might be that these are borderline sources with energies in the order of $10^{52}$ erg and show hybrid properties between BdHN I and II. However, this situation is new for us and needs further analysis and simulations to arrive at a definite answer.

%%%%%%%%%%%%%%%%%%%%%%%%%%%%%%%%%%%%%%%%%%%%%%%%%%%%%%%%%%%
\subsection{The long-short GRB connection}\label{sec:6b}
%%%%%%%%%%%%%%%%%%%%%%%%%%%%%%%%%%%%%%%%%%%%%%%%%%%%%%%%%%%

The above classification implies a unique prediction of this scenario: BdHNe I lead to NS-BH and BdHN II to NS-NS if the binary holds bound in the cataclysmic event. The BdHN III most likely unbinds the progenitor binary. Gravitational-wave emission drives the bound compact-object binaries to merge, leading to short GRBs \citep{2015PhRvL.115w1102F,2019ApJ...871...14B,2022PhRvD.106h3002B,2023Univ....9..332B,2024ApJ...966..219B}. The above long-short GRB connection is a unique prediction of the BdHN scenario with verifiable observational consequences \citep{2024arXiv240115702B}.

\begin{figure*}
    \centering
    \includegraphics[width=0.8\linewidth,clip]{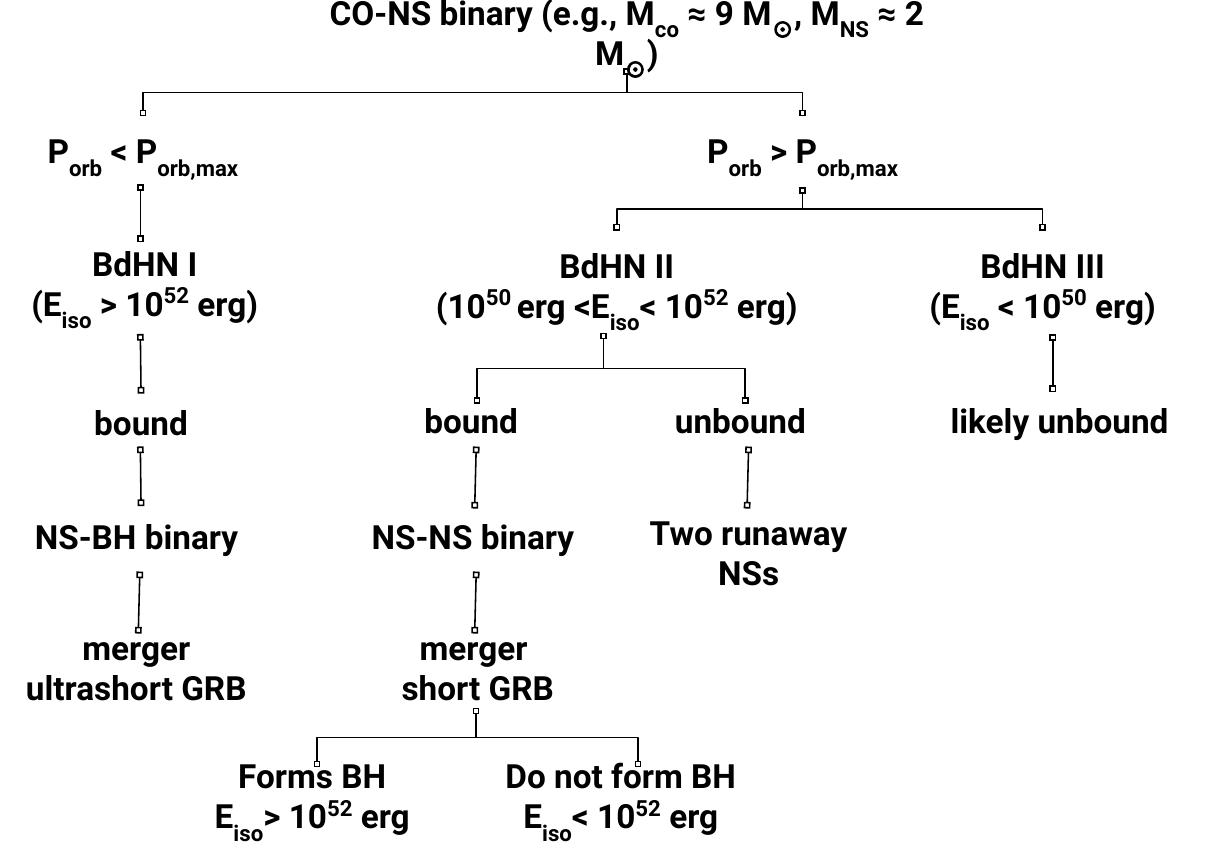}
    \caption{{Scheme of the BdHN I, II, and III scenarios, depending upon the pre-SN CO-NS orbital period. We recall that $P_{\rm orb,max}$ is the maximum orbital period for the NS companion to reach an instability point of BH formation, given CO and NS masses and SN kinetic energy.}}
    \label{fig:BdHNscheme}
\end{figure*}

If the system remains bound, the typical outcome of a BdHN I is an NS-BH, and of a BdHN II, an NS-NS. Gravitational-wave emission leads the NS-BH and NS-NS to merge, with the likely consequent emission of a short GRB \citep{2015PhRvL.115w1102F,2023Univ....9..332B,2024ApJ...966..219B}. Recent numerical simulations of the BdHN scenario for various binary parameters show a wide range of merger timescales $\sim 10^4$--$10^9$ yr \citep{2024arXiv240115702B}. The rapidly merging {(e.g., tens of kyr timescale)} binaries are those of short orbital periods (e.g., of a few minutes), so they are NS-BH. The wider binaries are NS-NS {and lead to longer merger times}. The fact that {the mass loss in} BdHNe should unbind a considerable amount of binaries {(see \citealp{2024arXiv240115702B}, for the latest simulations)} and the broad range of merger times is essential to explain the lower observed rate of short GRBs relative to that of long GRBs \citep[see, e.g.,][]{2016ApJ...832..136R, 2018ApJ...859...30R} and their shifted redshift distributions \citep[see, e.g.,][]{2014MNRAS.444L..58V, 2024ApJ...966..219B}. {Figure \ref{fig:BdHNscheme} shows a scheme of the above scenarios predicted by the BdHN model.}

Section \ref{sec:5c} has shown the possibility of forming BH-BH systems in some BdHN II if the $\nu$NS and the NS do not collapse by accretion but reach a final mass larger than the maximum mass allowed for the non-rotating configuration, i.e., if they end the accretion process in the NS supramassive region. This happens for soft EOS as the EOS with the eL$3\omega\rho$ parameterization with matter formed by a mix of nucleons and hyperons. These stars may reach secular instability by losing angular momentum by magnetic torque if the magnetic field is not buried by accretion. We showed they could form a BH-BH or a BH-NS (if only one of them is supramassive) in a timescale shorter than the merger timescale by gravitational-wave emission. Following the BH formation, in a timescale of tens of kyr, the gravitational-wave emission leads the BH-BH or BH-NS system to merge.\\

%%%%%%%%%%%%%%%%%%%%%%%%%%%%%%%%%%%%%%%%%%%%%%%%%%%%%%%%%%%
%%%%%%%%%%%%%%%%%%%%%%%%%%%%%%%%%%%%%%%%%%%%%%%%%%%%%%%%%%%
\section{Conclusions}\label{sec:7}
%%%%%%%%%%%%%%%%%%%%%%%%%%%%%%%%%%%%%%%%%%%%%%%%%%%%%%%%%%%
%%%%%%%%%%%%%%%%%%%%%%%%%%%%%%%%%%%%%%%%%%%%%%%%%%%%%%%%%%%

In this paper, we have followed the evolution of the $\nu$NS and its NS companion during their accretion of SN ejecta in a BdHN event leading to a long GRB. We have aimed to determine the condition under which these stars evolve into unstable configurations, reaching either the secular instability or the mass-shedding limit, collapsing into a {rotating, Kerr} BH. We assume the stars evolve through stable configurations as they accrete baryonic mass and angular momentum from the SN ejecta. The accretion rate onto the NSs is obtained from 3D SPH simulations using the adapted {\sc SNSPH code} of Los Alamos National Laboratory~\citep{SNSPH06}. To perform the evolution of the NS structure, we adapted the {\sc RNS code} \citep{1995ApJ...444..306S} to compute uniformly rotating NS configurations and used different parameterizations for the EOS, eL$3\omega\rho$, and NL$3\omega\rho$ both for cold and hot matter, with varying compositions including nucleons only and hyperons mixed with nucleons.

Our findings indicate that for low-mass progenitors of the carbon-oxygen (CO) star, the mass accretion rate onto the binary stars is insufficient to push them toward an unstable point. This holds, for instance, in the case of the CO progenitor with $M_{\rm ZAMS}=15M_\odot$. On the other hand, for the progenitors of the CO star with $M_{\rm ZAMS}=30 M_\odot$, the configurations reach the mass-shedding limit under the same conditions regardless of the assumed EOS. This happens for configurations with $j_{\rm NS, 0}>1.0$ and efficient angular momentum accretion. The secular instability limit is reached when the EOS with the eL$3\omega\rho$ parameterization is used for matter with a mix of nucleons and hyperons and configuration with $j_{\rm NS, 0}<1.0$. This is the softer EOS we have considered in our study.

The time required to collapse can be as short as $10$ s for rapidly rotating initial stars reaching the mass-shedding limit or as short as $50$ s for slowly rotating initial stars reaching the secular instability limit and modeled by a soft EOS. As expected, this time increases with the binary period of the CO-NS system. {There is the possibility that the appearance of the $\nu$NS at times of the order of $100$ s, e.g., owing to the second accretion peak onto it (see, e.g., simulation with an orbital period of about $10$ min in the left panel of Fig. \ref{fig:Mdots_Porb} and discussion in \citealp{2022PhRvD.106h3002B}), and the BH formation at comparable times (see left panel of Fig. \ref{fig:CollapseTime}), could have been individuated in three sources: GRB 221009A, GRB 221001A, and GRB 160625B  (Ruffini et al., plenary talk at the 17th Marcel Grossmann Meeting 2024, article in preparation). 
}

Summarizing, we confirm results from previous simulations \citep{2019ApJ...871...14B,2022PhRvD.106h3002B} that short-period CO-NS binaries can lead to BH formation in an orbit timescale. {In this work, we showed} the effect of the NS angular momentum: the time to {reach the conditions for gravitational collapse} shortens if the NS has non-zero angular momentum before the accretion process. It can become as short as tens of seconds for a rapidly rotating NS before the SN explosion triggers the GRB event. {This present approach is based on the simplest BdHN model based on a core collapse in the CO star of about $10 M_\odot$, forming the $\nu$NS and the SN, in the presence of an NS companion. However, the quality of data from the early phases of BdHNe, particularly identifying the SN-rise \citep{2024arXiv240508231R}, opens the possibility to explore alternative scenarios for the $\nu$NS and the SN formation (Ruffini et al., in preparation).}

All the above results provide an additional step toward a comprehensive understanding of binary stellar evolution, starting from binaries of main-sequence massive stars to intermediate stages like binary X-ray sources, whose further evolution leads to the most powerful transients in the Universe, long GRBs, and finally to compact-star binaries merging producing short GRBs.

%%%%%%%%%%%%%%%%%%%%%%%%%%%%%%%%%%%%%%%%%%%%%%%%%%%%
\begin{acknowledgments}
%%%%%%%%%%%%%%%%%%%%%%%%%%%%%%%%%%%%%%%%%%%%%%%%%%%%

L.M.B. is supported by the Vicerrector\'ia de Investigación y Extensi\'on - Universidad Industrial de Santander Postdoctoral Fellowship Program No.  2023000359. D.P.M. is partially supported by project INCT-FNA Proc. No. 464898/2014-5 and  Conselho Nacional de Desenvolvimento Científico e Tecnológico (CNPq/Brazil) under grant 303490/2021-7. C.P. acknowledges support from FCT (Fundação para a Ciência e a Tecnologia, I.P, Portugal) under Projects  UIDP/\-04564/\-2020.  UIDB/\-04564/\-2020 and 2022.06460.PTDC. 
\end{acknowledgments}

%\bibliographystyle{aasjournal}
%\bibliography{reference}

\end{document}